\documentclass[letter,12pt]{article}

\usepackage{jheppub}

\usepackage{float, array, xspace, amscd, amsthm}
\usepackage{fancyhdr}
\usepackage{longtable}

\usepackage{amsmath}
\usepackage{slashed}
\usepackage{microtype}
\usepackage{tabularx}
\usepackage{booktabs}

\makeatletter
\g@addto@macro\bfseries{\boldmath} %
\makeatother

\usepackage[dvipsnames]{xcolor}

\let\originalleft\left
\let\originalright\right
\renewcommand{\left}{\mathopen{}\mathclose\bgroup\originalleft} %
\renewcommand{\right}{\aftergroup\egroup\originalright}

\usepackage{amssymb}
\usepackage{amsfonts}
\usepackage{mathrsfs}
\usepackage{graphicx}
\usepackage{mathtools}
\usepackage{slashed}
\usepackage{amsthm}
\usepackage{amscd}
\usepackage{epsfig}                    %
\usepackage[matrix,arrow]{xy}          %
\usepackage{xspace}                    %
\usepackage{stmaryrd}                  %
\SetSymbolFont{stmry}{bold}{U}{stmry}{m}{n} %
\usepackage{slashed}
\usepackage{appendix,graphicx}
\usepackage{enumitem}
\usepackage{hyperref}
\usepackage{tensor}
\usepackage{units} %
\usepackage{tikz}
\usepackage{tikz-cd}

\newcommand{\dd}{{\rm d}}

\newcommand{\End}{\text{End}}
\newcommand{\tr}{\text{tr}}

\newcommand{\DD}{\bar{\mathcal{D}}}

\newcommand{\cN}{\mathcal{N}}

\def\bea#1\eea{\begin{align}#1\end{align}}

\usepackage{color}

\theoremstyle{definition}

\global\long\def\op#1{\operatorname{#1}}%

\newcommand{\ii}{\mathrm{i}}
\newcommand{\ee}{\mathrm{e}}

\newcommand{\der}{\partial}
\newcommand{\del}{\partial}
\newcommand{\delb}{\bar{\partial}}

\newcommand{\bbR}{\mathbb{R}}
\newcommand{\bbC}{\mathbb{C}}

\global\long\def\eqspace{\mathrel{\phantom{{=}}{}}}%

\DeclareMathOperator{\SU}{\mathit{SU}}

\DeclareMathOperator{\SO}{\mathit{SO}}

\newcommand{\rep}[1]{\boldsymbol{#1}}
\newcommand{\repb}[1]{\bar{\boldsymbol{#1}}}

\DeclareMathOperator{\ch}{ch}

\newcommand{\Pcomplex}{\text{P}}

\newcommand{\LC}{\nabla}

\newcommand{\inn}{\mathbin{\lrcorner}}

\newcommand{\ra}{\rightarrow}

\newcommand{\A}{\mathcal{A}}

\makeatletter
\gdef\@fpheader{~} %
\makeatother

\title{Holomorphic supergravity in ten dimensions and anomaly cancellation}
\author[a]{Anthony Ashmore,}
\emailAdd{aashmore@skidmore.edu}
\author[b]{Javier José Murgas Ibarra,}
\emailAdd{javier.j.murgasibarra@uis.no}
\author[c]{Charles Strickland-Constable,}
\emailAdd{c.strickland-constable@herts.ac.uk}
\author[b]{Eirik Eik Svanes}
\emailAdd{eirik.e.svanes@uis.no}

\affiliation[a]{Department of Physics, Skidmore College, Saratoga Springs, NY 12866, United States}

\affiliation[b]{Department of Mathematics and Physics, Faculty of Science and Technology,
University of Stavanger, N-4036, Stavanger, Norway}

\affiliation[c]{Department of Physics, Astronomy and Mathematics,
University of Hertfordshire, College Lane, Hatfield, AL10 9AB, United Kingdom}

\abstract{We formulate a ten-dimensional version of Kodaira--Spencer gravity on a Calabi--Yau five-fold that reproduces the classical Maurer--Cartan equation governing supersymmetric heterotic moduli. Quantising this theory's quadratic fluctuations, we show that its one-loop partition function simplifies to products of holomorphic Ray--Singer torsions and exhibits an anomaly that factorises as in $SO(32)$ and $E_8\times E_8$ supergravity. Based on this, we conjecture that this theory is the $SU(5)$-twisted version of ten-dimensional $N=1$ supergravity coupled to Yang--Mills and show that it is related to the type I Kodaira--Spencer theory of Costello--Li via a non-local field redefinition. The counter-terms needed to cancel the anomaly and retain gauge invariance for the one-loop effective theory reconstruct the differential of a recently discovered double-extension complex. This complex has non-tensorial extension classes and its first cohomology counts the infinitesimal moduli of heterotic compactifications modulo $\mathcal{O}(\alpha'^2)$ corrections.}

\dedicated{Dedicated to the memory of our dear friend and colleague,\\ Paul Francis de Medeiros, who sadly passed away in March 2025.}
 
\begin{document} 

\maketitle
\flushbottom

\section{Introduction}

Recently, there has been much progress in understanding both classical and quantum aspects of geometries resulting from compactifications of the heterotic string. Due to their torsional nature, which in even dimensions implies the geometry is non-K\"ahler, many tools of complex and algebraic geometry are no longer applicable. However, building on the moduli analysis of \cite{Anderson:2010mh, Anderson:2011cza, Anderson:2011ty}, inspired by Atiyah~\cite{Atiyah:1955}, a differential and a cohomology for counting infinitesimal moduli of these geometries was recently written down~\cite{Anderson:2014xha, delaOssa:2014cia}.\footnote{See also \cite{Garcia-Fernandez:2015hja} where the moduli problem was investigated by means of generalised geometry and shown to be elliptic, implying a finite-dimensional moduli space. For further investigations into classical (even-dimensional) heterotic moduli, see \cite{Candelas:2016usb, Candelas:2018lib, Garcia-Fernandez:2018emx, Ashmore:2018ybe, Ashmore:2019rkx, McOrist:2019mxh, Garcia-Fernandez:2020awc, 
McOrist:2021dnd, McOrist:2024glz}.}

Quantum aspects of heterotic moduli are interesting both from a physical perspective, in terms of topological quantum field theory, and a mathematical perspective, in terms of topological invariants and enumerative geometry. Indeed, these perspectives are linked, as knot invariants are known to correspond to correlation functions in Chern--Simons theory~\cite{Witten:1988hf}, while Donaldson--Thomas invariants can be calculated using holomorphic Chern--Simons theory~\cite{donaldson1998gauge, thomas1997gauge}. A theory capturing the complete moduli of heterotic geometries would not only be of physical importance, but would also be useful for enumerative geometry, particularly for classifying and counting non-K\"ahler geometries. An incomplete proposal for such a theory was put forward in \cite{Ashmore:2018ybe}, with the theory's one-loop partition function computed, and its topological properties and anomalies investigated in \cite{Ashmore:2023vji}. Still, many questions remain. How does the theory embed in ten-dimensional supergravity? What is the quantum theory for truly non-K\"ahler backgrounds?\footnote{The physical and geometric anomalies of the partition function of \cite{Ashmore:2023vji} was computed assuming a K\"ahler Calabi--Yau background.} What are the higher-loop correlation functions, and, relatedly, is there a holomorphic anomaly equation for them?\footnote{Quantum and invariant properties of the Hull--Strominger system have also been investigated in the mathematics literature. The interested reader is referred to \cite{Alvarez-Consul:2020hbl, Alvarez-Consul:2023zon, Garcia-Fernandez:2023nil, Tellez-Dominguez:2023wwr, dominguez2025moduli} with references therein.}

In looking to make the connection to ten-dimensional supergravity, it is natural to consider supersymmetric twists, a subject with a long history going back to~\cite{Witten:1990}. 
The twist of Euclidean super Yang--Mills theory on a Calabi--Yau three-fold was shown to give rise to holomorphic Chern--Simons theory in~\cite{Figueroa-OFarrill:1997dpz} (see also~\cite{Baulieu:1997jx,Blau:1997pp} which discuss this theory). 
Following this, the $SU(5)$ twist of ten-dimensional super Yang--Mills theory was found to be given by a ten-dimensional holomorphic Chern--Simons theory~\cite{Baulieu:2010ch}, with the field in the Chern--Simons term given by a fermionic $(0,2)$-form. Classically, holomorphic Chern--Simons theory is quasi-topological in the sense that the classical action depends on the data of the complex structure of the manifold and not on the metric. This is the sense in which we use the term quasi-topological in this article. However, anomalies can lead to a metric dependence in the quantum theory, which can be seen at the level of the one-loop partition function (see e.g.~\cite{Ashmore:2023vji}). 

More recently, a procedure for twisting supergravity theories was proposed by Costello and Li~\cite{Costello:2016mgj}, with much work devoted to investigating the resulting theories~\cite{Saberi:2021weg,Eager:2021ufo,Costello:2021kiv,Raghavendran:2021qbh,Hahner:2023kts}, often using the pure spinor superfield formalism.  In a similar fashion to the Yang--Mills case, it is believed that the $\SU(5)$ twist of supergravity in ten-dimensions will take a similar holomorphic/quasi-topological form. The study of the heterotic moduli problem has given hints towards what this form might be. In \cite{Ashmore:2018ybe} the superpotential of six-dimensional heterotic compactifications \cite{Becker:2003yv, LopesCardoso:2003dvb, Gurrieri:2004dt} was used to derive a cubic Chern--Simons-like theory for the geometric deformation parameters, with the resulting equations of motion giving the Maurer--Cartan equation for the heterotic moduli problem. Moreover, this theory reduces to ordinary holomorphic Chern--Simons theory when the geometric deformations (metric, complex structure and Kalb--Ramond field) are turned off, thus matching the twisted gauge theory expected when one twists supergravity coupled to Yang--Mills multiplets. The theory has a direct generalisation to other complex dimensions, such as for a complex five-fold $X$ with an $SU(5)$ structure. As we will describe, the classical BV-action for these theories takes the form 
\begin{equation}
\label{eq:BVaction0}
    S=\int_X\left(\langle y,\bar D y\rangle-\tfrac{1}{3}\langle y,[y,y]\rangle\right)\wedge\Omega\:,
\end{equation}
where $\Omega$ is the holomorphic top-form, $y\in \Pcomplex^{0,\bullet}_\sim$ is the BV-field, and $\Pcomplex^{0,\bullet}_\sim$ is a certain complex to be defined in detail below. Note that similar holomorphic theories have also appeared in studies of holomorphic BF-type field theories and in the context of twisted holography, see \cite{Giusto:2012jm, Costello:2015xsa, Costello:2019jsy, Costello:2021kiv, Bittleston:2024efo, Fernandez:2024qnu} and references therein. It should also be noted that the theory \eqref{eq:BVaction0} is well defined for more general torsional Hull--Strominger backgrounds with fluxes turned on.

It is natural to conjecture that the theory \eqref{eq:BVaction0} is twisted ten-dimensional supergravity coupled to Yang--Mills. A full proof of this for the non-linear theory is beyond the scope of the present paper -- here instead we test the conjecture by restricting to the quadratic theory. We compute the one-loop partition function and corresponding anomalies, assuming a Calabi--Yau background. We find the anomaly polynomials to be precisely those of ten-dimensional heterotic supergravity, in support of the conjecture. Furthermore, the counter-terms added to cancel the anomaly and retain a gauge invariant one-loop effective action give rise to the differential $\bar D$ of a recently discovered double extension complex~\cite{McOrist:2021dnd, McOrist:2024zdz, Chisamanga:2024xbm, deLazari:2024zkg, Ibarra:2024ntg}. Modulo ${\cal O}(\alpha'^2)$ terms, the first cohomology of this complex counts the infinitesimal deformations of heterotic solutions. In further support of our conjecture, the theory \eqref{eq:BVaction0} can be linked to the type I topological string of \cite{Costello:2015xsa, Costello:2019jsy, Costello:2021kiv} via a non-local field redefinition.

The paper is organised as follows. In Section \ref{sec:10d-theory}, we define the BV double complex for our linearised ten-dimensional Kodaira--Spencer theory, write the resulting quadratic master action, and show that the one-loop partition function of this theory collapses to a product of holomorphic Ray--Singer torsions. In Section \ref{sec:cubic_theory}, we focus on the moduli subsector and give a cubic Chern--Simons action whose equations of motion reproduce the Maurer--Cartan equation governing the moduli of the ten-dimensional supersymmetric background. We show that the one-loop partition function of this theory matches that of Section \ref{sec:10d-theory} and describe how the theory connects with the type I topological string. In Section \ref{sec:anomalies}, we derive the anomaly polynomial of our linearised ten-dimensional Kodaira--Spencer theory. Specialising to the gauge groups $SO(32)$ and $E_8\times E_8$, we find that the anomaly factorises. We then propose four ways to cancel the anomaly using: 1) non-local counter-terms; 2) a Green--Schwarz-like mechanism; 3) local counter-terms from deformed instanton equations; 4) non-global counter-terms. We conclude in Section \ref{sec:discussion} with some discussion and directions for further study.

\section{Linearised KS theory from ten-dimensional supergravity}
\label{sec:10d-theory}

The Costello--Li approach to supergravity suggests that twists of ten-dimensional $\mathcal{N}=1$ supergravity theories should produce a holomorphic or quasi-topological field theory on a complex five-fold with an $SU(5)$ structure. In this picture, the twisted theory is not the full physical theory, but a protected subsector that should still encode meaningful data, including a holomorphic deformation theory for the twisted background and the associated one-loop anomalies. One therefore expects the presence of a holomorphic moduli theory that plays a role analogous to holomorphic Chern--Simons theory for twisted gauge theory and to Kodaira--Spencer/BCOV theory for the closed-string sector in lower dimensions.

Motivated by this expectation, we start from the six-dimensional complex introduced in \cite{delaOssa:2014cia} and used to define a heterotic Kodaira--Spencer theory in \cite{Ashmore:2023vji}. We then formally generalise this complex to a ten-dimensional holomorphic theory on a complex five-fold. The resulting Maurer--Cartan equation has the same schematic structure as the heterotic deformation problem, now packaged in ten-dimensional language, and reduces to holomorphic Chern--Simons theory when the geometric fields are switched off. Having identified this classical holomorphic theory, we pass to a BV description in order to quantise it systematically: the resulting BV complex organises fields, ghosts, and antifields so that gauge symmetries are treated consistently and the path integral is well defined. This BV formulation is essential for our main tests of the proposal, namely the computation of the one-loop partition function and the calculation and cancellation of anomalies.

We begin by defining a linearised ten-dimensional theory as a straightforward analogue of the six-dimensional theory of \cite{Ashmore:2023vji}. 
In fact, what we do in this section is not quite the heterotic theory, so we first clarify this point. 
Let us recall that the full heterotic supergravity theory has a Bianchi identity for the three-form flux
\begin{equation} \label{eq:het-Bianchi}
     \dd H=\frac{\alpha'}{4}\bigl(\tr(F\wedge F)-\tr(R^+\wedge R^+)\bigr)\: .
\end{equation}
featuring the higher-derivative Riemann-squared term. (This term in fact features the Riemann tensor for the connection $\LC^+ = \LC + \frac12 H$~\cite{Bergshoeff:1989de}.) 
In formulating the theory described in this section, we essentially drop this term leaving the pure $\cN = 1$ supergravity theory coupled to Yang--Mills multiplets.\footnote{If one includes additional gauge fields, comprising an additional tangent bundle connection, and then examines solutions where this connection is set to be $\LC^+$, then the solutions of this augmented pure supergravity can be seen as containing those of the heterotic theory. 
However, in general there would also be solutions where the additional connection was not set to $\LC^+$, which would not be physically relevant. We discuss this further below.} 
The theory we construct in this section can thus be seen as a Kodaira--Spencer theory coming from two-derivative classical ten-dimensional supergravity. 
We will see how the Riemann-squared term arises from cancelling anomalies in this theory, in a similar way to how it appears from anomaly cancellation in heterotic supergravity. 

Furthermore, this theory really comes from a Wick-rotated (complexified) supergravity in Euclidean signature. For the case of one vector multiplet, this could be seen as a consistent truncation of eleven-dimensional supergravity on the time direction, in a similar way to how Einstein--Maxwell theory is found by reduction in \cite{Bergshoeff:1981um}. 
Note that the supersymmetric solutions of this theory will be constrained by no-go theorems, which generally prevent compact solutions with non-trivial fluxes. 
More non-trivial solutions are permitted once one includes the Riemann-squared terms coming from anomaly cancellation. 

To avoid confusion, let us also expand on the point that it is common practice, especially in the mathematics literature on solutions of the Hull--Strominger system, to include a tangent bundle connection as part of the gauge degrees of freedom. This has the effect of promoting the Bianchi identity to 
\begin{equation} 
    \ii \, \partial\bar\partial\omega=\frac{\alpha'}{4}\bigl(\tr(F\wedge F)-\tr(R\wedge R)\bigr)\:,
\end{equation}
where $R$ is the curvature of the tangent bundle connection. This choice, however, introduces extra, spurious degrees of freedom associated to the tangent bundle connection which are not present in heterotic string theory (where the connection is defined by the physical metric). The resulting space of solutions does however contain the physical ones: those for which the spurious tangent bundle connection takes the correct physical value $\nabla^+$. In this paper, we will refer to the space of deformations of these solutions inside the enlarged space as the moduli space of the Hull--Strominger system (plus spurious degrees of freedom). In general, this moduli space is not the physical one as it will contain deformations that break the condition that the additional tangent bundle connection is $\nabla^+$. 

One might hope there are interesting non-K\"ahler solutions without including the $R^2$ term in~\eqref{eq:het-Bianchi}, however an examination of the remaining supersymmetry conditions shows there are no such solutions. In particular, the supersymmetry conditions for the background include the Yang--Mills equation $\omega^4\wedge F=0$. Unfortunately, the only solutions to this and the Bianchi identity without Riemann-squared terms (Equation \eqref{eq:F_Bianchi} below) for compact geometries are flat bundles on Calabi--Yau manifolds \cite{Garcia-Fernandez:2023vah}.

\subsection{The BV complex from supergravity}
\label{sec:10d-BV-complex}

We start by discussing the complex of de la Ossa and Svanes~\cite{delaOssa:2014cia}, though generalised to ten dimensions. 
As mentioned above, we in fact consider the version of this complex that features the gauge fields of the pure (Euclidean) supergravity theory, but not the tangent bundle connection which would give rise to the Riemann squared term in the Bianchi identity. 
This complex computes the moduli of supersymmetric solutions to this Euclidean theory.\footnote{One way to see this is to use the generalised geometry formalism, as in~\cite{Ashmore:2019rkx,Kupka:2024vrd,Kupka:2024xur}, where the supersymmetry conditions become the statement that the $\SU(5)\times\SO(10+n)$ structure defined by the Killing spinor has vanishing intrinsic torsion.} 
Let $X$ be a complex five-fold. We define a holomorphic extension bundle 
\begin{equation}
    Q \simeq  \Lambda^{1,0} \oplus \End(V)  \oplus T^{1,0} \:,
\end{equation}
where $T^{1,0}\equiv T^{1,0}X$ is the holomorphic tangent bundle of $X$, $\Lambda^{1,0} \equiv T^{*1,0}X$, and $V$ is a holomorphic vector bundle over $X$. There is then a holomorphic differential on $\Omega^{0,\bullet}(Q)$ given by
\begin{equation}
\label{eq:barD1}
	\bar{D} = \begin{pmatrix}
\delb & \,\alpha'\mathcal{F}^*\, &  {\cal T}\\
0 & \delb_\A & \mathcal{F}\\
0 & 0 & \delb 
\end{pmatrix}\:.
\end{equation}
where the various extension map components of $\bar D$ act as
\begin{align}
    \mathcal{F}(\mu)&=F_{a}\wedge\mu^a \:,\\
    \left(\mathcal{F}^*(\alpha)\right)_a&=\tr(F_{a}\wedge\alpha)\:,\\
    \left(\mathcal{T}(\mu)\right)_a&=-2\,T_{ab}\wedge\mu^b\:,
\end{align}
where $F_a=F_{a\bar b}\dd z^{\bar b}$ is defined from a curvature two-form $F$ for $V$, and $T_{ab}=T_{ab\bar c}\dd z^{\bar c}$ is given by the $(2,1)$-component of the three-form flux $H$. For supergravity solutions, this is given by $T=\ii\,\partial\omega$, where $\omega$ is the hermitian two-form of the (in general non-K\"ahler) geometry on $X$. The operator \eqref{eq:barD1} then squares to zero if and only if
\begin{equation}\label{eq:F_Bianchi}
    \ii \, \partial\bar\partial\omega=\frac{\alpha'}{4}\tr(F\wedge F)\:,
\end{equation}
which is the standard supergravity Bianchi identity.  

This kind of operator has appeared 
in the study of heterotic moduli problems~\cite{Anderson:2014xha, delaOssa:2014cia, Garcia-Fernandez:2015hja}, and generalised geometry and coupled instantons~\cite{Garcia-Fernandez:2021lyt, Garcia-Fernandez:2023vah, Garcia-Fernandez:2023nil, Silva:2024fvl, Garcia-Fernandez:2024ypl}.
As discussed above, these authors consider the gauge fields to include an additional tangent bundle connection, fixed to be $\nabla^+$, to promote~\eqref{eq:F_Bianchi} to the full Bianchi identity including the Riemann-squared terms, at the cost of introducing spurious degrees of freedom. 

In this work, we will take a different approach to including the Riemann-squared terms. When studying quantum aspects of the theory and anomalies in Section \ref{sec:anomalies}, we will choose a diagonal differential with vanishing fluxes for the classical theory, $\bar D=\delb$. Using the Green--Schwarz anomaly cancellation procedure, together with the requirement that the one-loop effective action is gauge invariant, we will then correct the differential back to an extension structure. This approach avoids the introduction of the spurious modes and allows for the anomaly to be calculated more straightforwardly. It does however render the differential $\bar D$ slightly unconventional in that its off-diagonal terms are non-tensorial. In particular, $\bar D$ is no longer the $(0,1)$-part of a connection on $Q$. For now, the operator $\bar D$ will refer to equation \eqref{eq:barD1} with the off-diagonal flux terms turned on.

In the physics literature, by considering F-term constraints, the first cohomology of the complex $(\Omega^{0,\bullet}(Q), \bar{D})$ was shown to compute the infinitesimal moduli space of a solution to the Hull--Strominger system (plus the spurious degrees of freedom) in~\cite{delaOssa:2014cia, Anderson:2014xha, delaOssa:2015maa}. This statement is true also for the ten-dimensional (Euclidean) version of the system. The moduli problem has also been investigated formally in the mathematics literature and connected to generalised geometry~\cite{Garcia-Fernandez:2015hja, Ashmore:2019rkx}. In particular, the moduli problem was shown to be elliptic in \cite{Garcia-Fernandez:2015hja}, giving rise to a finite-dimensional moduli space on compact geometries. 

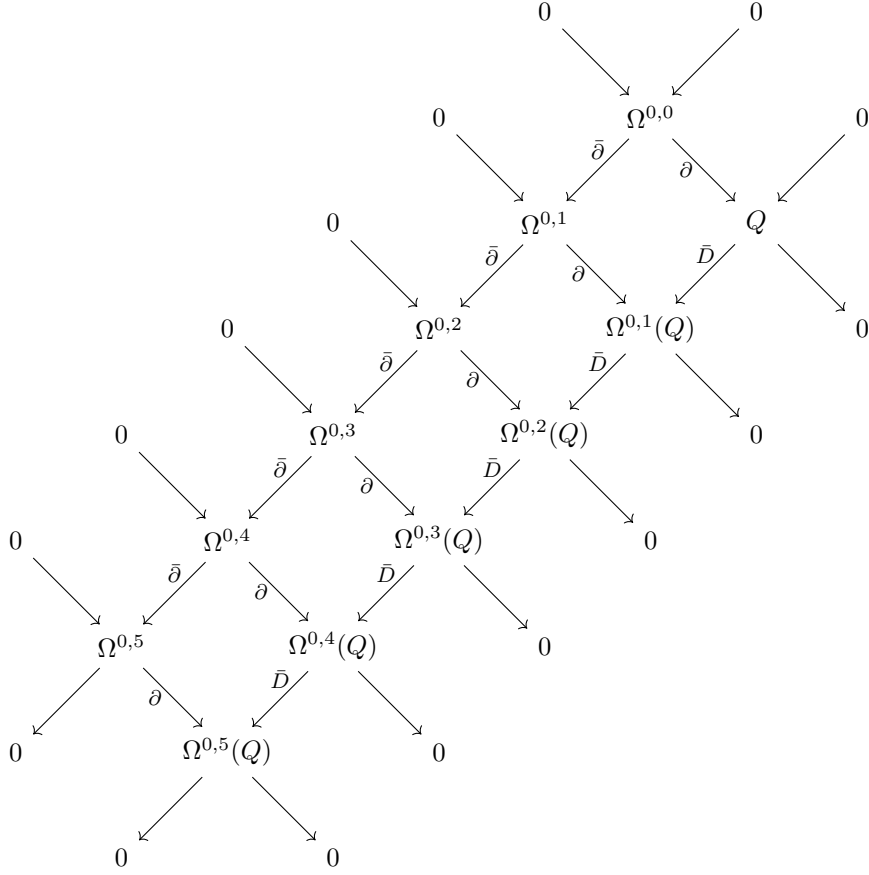
\begin{figure}
\begin{tikzpicture}[scale=1.4,baseline=(current bounding box.center),font=\footnotesize] %% centers equation number
%\node (z) at (-2.5,2) {}; 
%\node (h) at (-1.3,2) {0}; 
%\node (G) at (-7,0.5) {Ghost \#}; 
%\node (L0) at (-7,0) {3}; 
%\node (L1) at (-7,-1) {2}; 
%\node (L2) at (-7,-2) {1}; 
%\node (L3) at (-7,-3) {0}; 
%\node (L4) at (-7,-4) {-1}; 
%\node (L5) at (-7,-5) {-2}; 
%\node (L6) at (-7,-6) {-3}; 
%\node (L7) at (-7,-7) {-4}; 
%\node (L8) at (-7,-8) {}; 
%%%%%%
\node (0) at (-1,1) {0}; 
\node (1) at (-2,0) {0}; 
\node (2) at (-3,-1) {0}; 
\node (3) at (-4,-2) {0}; 
\node (4) at (-5,-3) {0}; 
\node (5) at (-6,-4) {0}; 
\node (6) at (-7,-5) {}; 
%\node (Az) at (-2.5,1) {$0$}; 
\node (Ah) at (1,1) {$0$}; 
\node (A0) at (0,0) {$\Omega^{0,0}$}; 
\node (A1) at (-1,-1) {$\Omega^{0,1}$}; 
\node (A2) at (-2,-2) {$\Omega^{0,2}$}; 
\node (A3) at (-3,-3) {$\Omega^{0,3}$}; 
\node (A4) at (-4,-4) {$\Omega^{0,4}$}; 
\node (A5) at (-5,-5) {$\Omega^{0,5}$}; 
\node (A6) at (-6,-6) {$0$}; 
%\node (Bz) at (-2.5,0) {$0$}; 
\node (Bh) at (2,0) {$0$}; 
\node (B0) at (1,-1) {$Q$}; 
\node (B1) at (0,-2) {$\Omega^{0,1}(Q)$}; 
\node (B2) at (-1,-3) {$\Omega^{0,2}(Q)$}; 
\node (B3) at (-2,-4) {$\Omega^{0,3}(Q)$}; 
\node (B4) at (-3,-5) {$\Omega^{0,4}(Q)$}; 
\node (B5) at (-4,-6) {$\Omega^{0,5}(Q)$}; 
\node (B6) at (-5,-7) {$0$}; 
%\node (Cz) at (-2.5,-1) {$0$}; 
%\node (Ch) at (3,-1) {$0$}; 
\node (C0) at (2,-2) {$0$}; 
\node (C1) at (1,-3) {$0$}; 
\node (C2) at (0,-4) {$0$}; 
\node (C3) at (-1,-5) {$0$}; 
\node (C4) at (-2,-6) {$0$}; 
\node (C5) at (-3,-7) {$0$}; 
%\node (C6) at (-4,-8) {$0$}; 
%\node (Dz) at (-2.5,-2) {}; 
%\node (Dh) at (-1.3,-2) {0}; 
%\node (D0) at (3,-3) {0}; 
%\node (D1) at (2,-4) {0}; 
%\node (D2) at (1,-5) {0}; 
%\node (D3) at (0,-6) {0}; 
%\node (D4) at (-1,-7) {0}; 
%\node (D5) at (-2,-8) {0}; 
%\node (D6) at (-3,-9) {}; 
\path[->,font=\scriptsize] 
%(Az) edge node[above]{$$} (Ah)
(Ah) edge node[above]{} (A0)
(A0) edge node[above]{$\bar\der$} (A1)
(A1) edge node[above]{$\bar\der$} (A2)
(A2) edge node[above]{$\bar\der$} (A3)
(A3) edge node[above]{$\bar\der$} (A4)
(A4) edge node[above]{$\bar\der$} (A5)
(A5) edge node[above]{} (A6)
%(Bz) edge node[above]{$$} (Bh)
(Bh) edge node[above]{} (B0)
(B0) edge node[above]{$\bar{D}$} (B1)
(B1) edge node[above]{$\bar{D}$} (B2)
(B2) edge node[above]{$\bar{D}$} (B3)
(B3) edge node[above]{$\bar{D}$} (B4)
(B4) edge node[above]{$\bar{D}$} (B5)
(B5) edge node[above]{} (B6)
%(Cz) edge node[above]{$$} (Ch)
%(Ch) edge node[above]{} (C0)
%(C0) edge node[above]{$\bar\der$} (C1)
%(C1) edge node[above]{$\bar\der$} (C2)
%(C2) edge node[above]{$\bar\der$} (C3)
%(C3) edge node[above]{$\bar\der$} (C4)
%(C4) edge node[above]{$\bar\der$} (C5)
%(C5) edge node[above]{} (C6)
%(h) edge node[left]{} (Ah)
(0) edge node[left]{} (A0)
(1) edge node[left]{} (A1)
(2) edge node[left]{} (A2)
(3) edge node[left]{} (A3)
(4) edge node[left]{} (A4)
(5) edge node[left]{} (A5)
%(Ah) edge node[left]{$\der$} (Bh)
(A0) edge node[left]{$\der$} (B0)
(A1) edge node[left]{$\der$} (B1)
(A2) edge node[left]{$\der$} (B2)
(A3) edge node[left]{$\der$} (B3)
(A4) edge node[left]{$\der$} (B4)
(A5) edge node[left]{$\der$} (B5)
%(Bh) edge node[left]{$\der$} (Ch)
(B0) edge node[left]{$$} (C0)
(B1) edge node[left]{$$} (C1)
(B2) edge node[left]{$$} (C2)
(B3) edge node[left]{$$} (C3)
(B4) edge node[left]{$$} (C4)
(B5) edge node[left]{$$} (C5);
%(Ch) edge node[left]{} (Dh)
%(C0) edge node[left]{} (D0)
%(C1) edge node[left]{} (D1)
%(C2) edge node[left]{} (D2)
%(C3) edge node[left]{} (D3)
%(C4) edge node[left]{} (D4)
%(C5) edge node[left]{} (D5);
\end{tikzpicture}
\caption{The double complex encoding the BV description of the complex $(\Omega^{0,\bullet}(Q), \bar{D})$ which computes the infinitesimal moduli space of the Hull--Strominger system (plus spurious degrees of freedom) on a complex five-manifold. The original complex is promoted to a double complex to account for gauge symmetries.}
\label{fig:10d-Q-complex-BRST}
\end{figure}

Here, we wish to construct a field theory from this complex via the BV formalism. This is complicated by the fact that there are additional gauge symmetries present in the construction. For example, the deformation of the background given by $(\mu,x,b) \in \Omega^{0,1}(T^{1,0}) \oplus \Omega^{1,1}_\bbC \oplus \Omega^{0,2}$ is gauge equivalent to that with $x \rightarrow x + \der \lambda$ for any $\lambda \in \Omega^{0,1}$. This leads us to extend the complex to a double complex, shown in Figure \ref{fig:10d-Q-complex-BRST}.
Ideally, we would then construct a quadratic field theory for this complex. Unfortunately, it does not possess a symplectic pairing of the type required by the BV formalism. 

\begin{figure}
\begin{tikzpicture}[scale=1.4,baseline=(current bounding box.center),font=\footnotesize] %% centers equation number
%\node (z) at (-2.5,2) {}; 
%\node (h) at (-1.3,2) {0}; 
\node (G) at (-7,0.5) {Ghost \#}; 
\node (L0) at (-7,0) {$3$}; 
\node (L1) at (-7,-1) {$2$}; 
\node (L2) at (-7,-2) {$1$}; 
\node (L3) at (-7,-3) {$0$}; 
\node (L4) at (-7,-4) {$-1$}; 
\node (L5) at (-7,-5) {$-2$}; 
\node (L6) at (-7,-6) {$-3$}; 
\node (L7) at (-7,-7) {$-4$}; 
\node (L8) at (-7,-8) {}; 
%%%%%%
\node (0) at (-1,1) {0}; 
\node (1) at (-2,0) {0}; 
\node (2) at (-3,-1) {0}; 
\node (3) at (-4,-2) {0}; 
\node (4) at (-5,-3) {0}; 
\node (5) at (-6,-4) {0}; 
\node (6) at (-7,-5) {}; 
%\node (Az) at (-2.5,1) {$0$}; 
\node (Ah) at (1,1) {$0$}; 
\node (A0) at (0,0) {$\Omega^{0,0}$}; 
\node (A1) at (-1,-1) {$\Omega^{0,1}$}; 
\node (A2) at (-2,-2) {$\Omega^{0,2}$}; 
\node (A3) at (-3,-3) {$\Omega^{0,3}$}; 
\node (A4) at (-4,-4) {$\Omega^{0,4}$}; 
\node (A5) at (-5,-5) {$\Omega^{0,5}$}; 
\node (A6) at (-6,-6) {$0$}; 
%\node (Bz) at (-2.5,0) {$0$}; 
\node (Bh) at (2,0) {$0$}; 
\node (B0) at (1,-1) {$Q$}; 
\node (B1) at (0,-2) {$\Omega^{0,1}(Q)$}; 
\node (B2) at (-1,-3) {$\Omega^{0,2}(Q)$}; 
\node (B3) at (-2,-4) {$\Omega^{0,3}(Q)$}; 
\node (B4) at (-3,-5) {$\Omega^{0,4}(Q)$}; 
\node (B5) at (-4,-6) {$\Omega^{0,5}(Q)$}; 
\node (B6) at (-5,-7) {$0$}; 
%\node (Cz) at (-2.5,-1) {$0$}; 
\node (Ch) at (3,-1) {$0$}; 
\node (C0) at (2,-2) {$\Omega^{5,0} $}; 
\node (C1) at (1,-3) {$\Omega^{5,1}$}; 
\node (C2) at (0,-4) {$\Omega^{5,2}$}; 
\node (C3) at (-1,-5) {$\Omega^{5,3}$}; 
\node (C4) at (-2,-6) {$\Omega^{5,4}$}; 
\node (C5) at (-3,-7) {$\Omega^{5,5}$}; 
\node (C6) at (-4,-8) {$0$}; 
%\node (Dz) at (-2.5,-2) {}; 
%\node (Dh) at (-1.3,-2) {0}; 
\node (D0) at (3,-3) {0}; 
\node (D1) at (2,-4) {0}; 
\node (D2) at (1,-5) {0}; 
\node (D3) at (0,-6) {0}; 
\node (D4) at (-1,-7) {0}; 
\node (D5) at (-2,-8) {0}; 
%\node (D6) at (-3,-9) {}; 
\path[->,font=\scriptsize] 
%(Az) edge node[above]{$$} (Ah)
(Ah) edge node[above]{} (A0)
(A0) edge node[above]{$\bar\der$} (A1)
(A1) edge node[above]{$\bar\der$} (A2)
(A2) edge node[above]{$\bar\der$} (A3)
(A3) edge node[above]{$\bar\der$} (A4)
(A4) edge node[above]{$\bar\der$} (A5)
(A5) edge node[above]{} (A6)
%(Bz) edge node[above]{$$} (Bh)
(Bh) edge node[above]{} (B0)
(B0) edge node[above]{$\bar{D}$} (B1)
(B1) edge node[above]{$\bar{D}$} (B2)
(B2) edge node[above]{$\bar{D}$} (B3)
(B3) edge node[above]{$\bar{D}$} (B4)
(B4) edge node[above]{$\bar{D}$} (B5)
(B5) edge node[above]{} (B6)
%(Cz) edge node[above]{$$} (Ch)
(Ch) edge node[above]{} (C0)
(C0) edge node[above]{$\bar\der$} (C1)
(C1) edge node[above]{$\bar\der$} (C2)
(C2) edge node[above]{$\bar\der$} (C3)
(C3) edge node[above]{$\bar\der$} (C4)
(C4) edge node[above]{$\bar\der$} (C5)
(C5) edge node[above]{} (C6)
%(h) edge node[left]{} (Ah)
(0) edge node[left]{} (A0)
(1) edge node[left]{} (A1)
(2) edge node[left]{} (A2)
(3) edge node[left]{} (A3)
(4) edge node[left]{} (A4)
(5) edge node[left]{} (A5)
%(Ah) edge node[left]{$\der$} (Bh)
(A0) edge node[left]{$\der$} (B0)
(A1) edge node[left]{$\der$} (B1)
(A2) edge node[left]{$\der$} (B2)
(A3) edge node[left]{$\der$} (B3)
(A4) edge node[left]{$\der$} (B4)
(A5) edge node[left]{$\der$} (B5)
%(Bh) edge node[left]{$\der$} (Ch)
(B0) edge node[left]{$\der$} (C0)
(B1) edge node[left]{$\der$} (C1)
(B2) edge node[left]{$\der$} (C2)
(B3) edge node[left]{$\der$} (C3)
(B4) edge node[left]{$\der$} (C4)
(B5) edge node[left]{$\der$} (C5)
%(Ch) edge node[left]{} (Dh)
(C0) edge node[left]{} (D0)
(C1) edge node[left]{} (D1)
(C2) edge node[left]{} (D2)
(C3) edge node[left]{} (D3)
(C4) edge node[left]{} (D4)
(C5) edge node[left]{} (D5);
\end{tikzpicture}
\caption{The double complex obtained by starting from Figure \ref{fig:10d-Q-complex-BRST} and adding a further diagonal. The ghost number of each space shown on the left. This BV complex is the ten-dimensional version of the complex of that appeared in \cite{Ashmore:2023vji} and, crucially, admits a symplectic pairing which can be used to define an action.}
\label{fig:10d-Q-complex}
\end{figure}
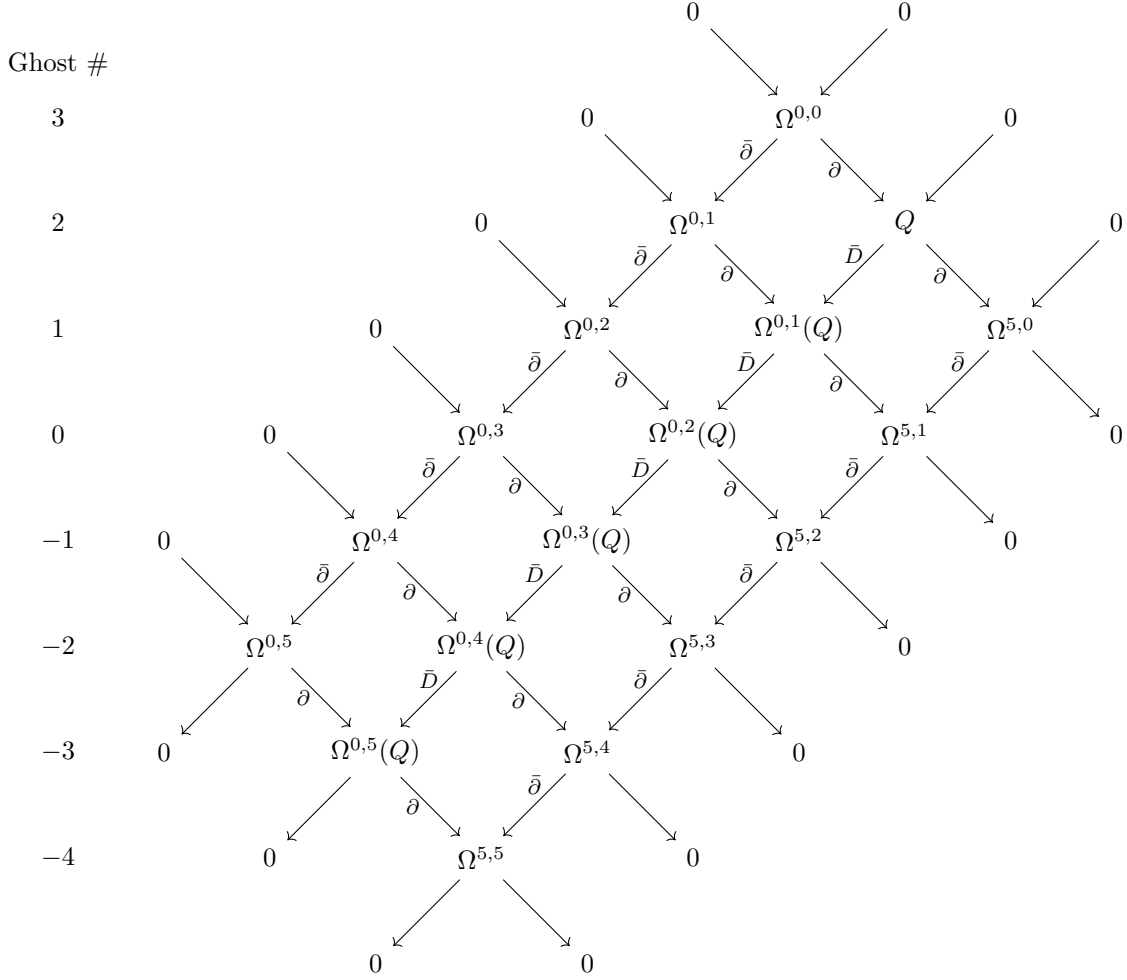

To fix this, we can construct a related complex which admits the required symplectic pairing by adding another diagonal corresponding to the dual spaces in the first diagonal (as was done in the generalised geometry context in~\cite{Kupka:2024rvl}). We then obtain a total differential that squares to zero by adding the maps $\der$, indicated below, between the sequence of new spaces. With this, we again have a double complex, which is simply the ten-dimensional version of the complex of \cite{Ashmore:2023vji}. The resulting double complex is shown in Figure \ref{fig:10d-Q-complex} with the ghost number of each space shown on the left.
Note that the bundle $Q$ can be written equivalently as
\begin{equation}
    Q \simeq \Lambda^{1,0} \oplus \End(V) \oplus T^{1,0} \simeq \Lambda^{1,0} \oplus \End(V) \oplus \Lambda^{4,0} \:,
\end{equation}
where we have used the isomorphism defined by the holomorphic volume form. The double complex in Figure \ref{fig:10d-Q-complex} has a symplectic pairing (or ``cyclic structure" in the $L_\infty$ algebra language) which pairs the spaces are diametrically opposite each other, taking the arrow between $\Omega^{0,2}(Q)$ and $\Omega^{0,3}(Q)$ to be the central point. For example, this pairs the terms $b^{0,2} \in \Omega^{0,2}$ and $c^{5,3} \in \Omega^{5,3}$ via the natural integration pairing
\begin{equation}
\label{eq:10d-pairing}
   \langle b^{0,2}, c^{5,3} \rangle = \int_X  b^{0,2} \wedge c^{5,3}\:. 
\end{equation}

\subsection{The classical master action}
The fields and anti-fields of linearised ten-dimensional Kodaira--Spencer theory in the BV formalism fill out a general element of the double complex in Figure \ref{fig:10d-Q-complex}. The ghost-number-zero fields (the ``physical'' fields) are fermionic, as are the other fields of even ghost number. The odd ghost number fields are bosonic. We denote elements of the various spaces that appear in the double complex by
\begin{equation}
   y^{0,\bullet} \in \Omega^{0,\bullet}(Q)\:, \qquad
   b^{0,\bullet} \in \Omega^{0,\bullet}\:, \qquad
   c^{5,\bullet} \in \Omega^{5,\bullet}\:.
\end{equation}
Note that the bosonic ghost field $y^{0,1}$ may be interpreted as infinitesimal fluctuations of the background geometry. The classical master action then takes the form\footnote{In this section of the paper, we will not pay close attention to the signs of the terms in the action. More details of these signs can be found in Section~\ref{sec:cubic_theory}.}
\begin{equation}
\label{eq:BVaction1}
   S = \langle \Phi , \dd_{\text{total}} \Phi \rangle 
   = \int_X \langle y, \bar{D} y \rangle \wedge \Omega 
   + \int_X b \wedge \bar\der c
   + \int_X (y \inn \der b) \wedge \Omega\:,
\end{equation}
where $\dd_{\text{total}}$ is the total differential  of the double complex of Figure  \ref{fig:10d-Q-complex}. We write fields in $\Omega^{0,\bullet}(Q)$ as
\begin{equation}
y=(x,\alpha,\mu)\quad\in\quad\Omega^{0,\bullet}(Q)=\Omega^{0,\bullet}(\Lambda^{1,0})\oplus\Omega^{0,\bullet}(\End(V))\oplus\Omega^{0,\bullet}(T^{1,0})\:.    
\end{equation}
The pairing on $\Omega^{0,\bullet}(Q)$, for $y_1\in\Omega^{0,p}(Q)$ and $y_2\in\Omega^{0,q}(Q)$, is given by
\begin{equation}
\langle y_1,y_2\rangle=(\mu_1)^a\wedge (x_2)_a+(x_1)_a\wedge(\mu_2)^a+\tr(\alpha_1\wedge\alpha_2)\:,
\end{equation}
where the indices on $x$ and $\mu$ labels the $\Lambda^{1,0}$ and $T^{1,0}$ components respectively.

The total pairing that appears in the action satisfies integration by parts with respect to the total differential $\dd_{\text{total}}$. The action \eqref{eq:BVaction1} is then invariant under the following gauge symmetries:
\begin{equation}\label{eq:gauge_y}
    \delta y=\bar D\gamma +\partial\beta\:,\qquad\delta b=\delb\beta\:,\qquad\delta c=\del(v\inn\Omega)+\delb\kappa\:,
\end{equation}
where $\gamma=(\xi,\epsilon,v)\in\Omega^{0,\bullet}(Q)$, $\beta\in\Omega^{0,\bullet}$, and $\kappa\in\Omega^{5,\bullet}$. These transformations can equivalently be written using the total differential as
\begin{equation}
    \delta\Phi = \dd_\text{total} \Lambda \:,
\end{equation}
for some element $\Lambda$ of the double complex.

\subsection{The one-loop partition function}

The action \eqref{eq:BVaction1} describes quadratic fluctuations of the background in the BV formalism. We now want to quantise this theory and compute its one-loop partition function. As discussed in \cite{Ashmore:2023vji}, in the quadratic theory, one can remove the $y \inn \partial b$ term in the action \eqref{eq:BVaction1} using field redefinitions. This leaves us with a classical master action of the form
\begin{equation}
\begin{aligned}
\label{eq:BVaction}
   S &= \int_X \langle y, \bar{D} y \rangle \wedge \Omega 
   + \int_X b \wedge \bar\der c \\
   &= \int_X\left( \langle y^{0,2}, \bar{D} y^{0,2} \rangle
   +\langle y^{0,4}, \bar{D} y^{0,0} \rangle\right)\wedge\Omega
   + b^{0,3} \wedge \bar\der c^{5,1}
   + b^{0,1} \wedge \bar\der c^{5,3} 
   \quad \text{(fermions)} \\
   &\eqspace+\int_X\langle y^{0,3}, \bar{D} y^{0,1} \rangle\wedge\Omega
   + b^{0,4} \wedge \bar\der c^{5,0}
   + b^{0,2} \wedge \bar\der c^{5,2}
   + b^{0,0} \wedge \bar\der c^{5,4}.
    \quad \text{(bosons)}
\end{aligned}
\end{equation}
The action \eqref{eq:BVaction} then fixes the one-loop partition function of the theory in terms of products of determinants of the various differential operators:\footnote{As usual, a path integral over one commuting (bosonic) field with kernel $K$ results in $(\det K)^{-1/2}$, while one anticommuting (fermionic) field gives $(\det K)^{+1/2}$. With two independent fields the exponents double, so two bosons yield $(\det K)^{-1}$ and two fermions $(\det K)^{+1}$. The second line then follows from the equivalence $|\Delta_{\bar\partial}^{0,q}| = |\bar{\partial}^{0,q}|^2 |\bar{\partial}^{0,q-1}|^2$, and similarly for the $\bar{D}$-Laplacian.}
\begin{equation}
\begin{aligned}
\label{eq:PartFunc}
   |Z| &= |\bar{D}^{0,2}|^{+1/2}
   |\bar{D}^{0,1}|^{-1}
   |\bar{D}^{0,0}|^{+1}
   |\bar\der^{0,0}|^{-1}
   |\bar\der^{0,1}|^{+1}
   |\bar\der^{0,2}|^{-1}
   |\bar\der^{0,3}|^{+1}
   |\bar\der^{0,4}|^{-1} \\
   &= \left[ \prod_{p=0}^{5} |\Delta^{0,p}_{\bar{D}}|^{(-1)^{p+1} p} \right]^{1/4}
   \left[ \prod_{q=0}^{5} |\Delta^{0,q}_{\bar\der}|^{(-1)^{q+1} q} \right]^{-1/2} \\
   &= I_{\bar{D}} (Q)^{1/2} I(\Lambda^{0,0})^{-1} \:.
\end{aligned}
\end{equation}
Here, we have written the absolute value of the one-loop partition function as an alternating product of determinants, which can then be simplified to a product of holomorphic Ray--Singer torsions for $\Lambda^{0,0}$ and $Q$.\footnote{Recall that given a holomorphic bundle $V$ with hermitian metric $h$ on a complex $n$-fold, the holomorphic Ray--Singer torsion $I(V)$ of $V$ is defined to be 
\begin{equation}
    I(V)^2 = \prod_{p=0}^n |\Delta_{\bar\partial}^{0,p}|^{(-1)^{p+1}p}\:.
\end{equation}}

The above determinant formula for the partition function highlights that the quadratic theory depends only on the complex structure via the holomorphic Ray--Singer torsions. We will interpret this in terms of an $SU(5)$ twist in the sections to come.

Although Ray--Singer torsions have been computed in certain idealised scenarios, such as for odd-dimensional spheres~\cite{weng1996analytic}, an explicit computation of $|Z|$, and in particular the holomorphic Ray--Singer torsion $I_{\bar{D}} (Q)$ is far beyond the state of current technology. Rather than an explicit computation of $|Z|$, we will focus on whether or not the partition function is actually well defined. To examine this, we compute the anomaly of $|Z|$ in Section \ref{sec:anomalies}. As we shall see, the anomaly will match the Green--Schwarz anomaly of ten-dimensional supergravity, suggesting a strong connection between our theory here and supergravity.

\section{The cubic theory for heterotic moduli}\label{sec:cubic_theory}

In the previous section, we constructed a BV master action in which the bosonic ghost field $y^{0,1}$ should be interpreted as capturing a deformation of a bosonic supersymmetric background of 
Euclidean ten-dimensional supergravity (or of heterotic supergravity including the spurious degrees of freedom if one includes the additional tangent bundle connection) on a complex five-fold. In this section, we first isolate the moduli subsector and the quadratic part of the action for it. We then extend the quadratic action to a cubic Chern--Simons-like theory, the relevant parts of whose field equations reproduces the non-linear Maurer--Cartan equation for moduli of the supersymmetric solution 
(i.e.~the ten-dimensional analogue of the Hull--Strominger system (including spurious degrees of freedom) if one includes the additional tangent bundle connection), as discussed in~\cite{Ashmore:2018ybe} in the six-dimensional case.

The classical master action for the quadratic theory in \eqref{eq:BVaction1} naturally presents the fermionic fields $y^{0,2}$, $b^{0,3}$ and $c^{5,1}$ of ghost number zero as the ``physical'' fields. However, guided by the six-dimensional superpotential theory of \cite{Ashmore:2018ybe}, the bosonic ghost field $y^{0,1}=(x,\alpha,\mu)$ should be interpreted as geometric moduli fields, where $x\in\Omega^{0,1}(\Lambda^{1,0})$ give the (complexified) hermitian moduli, $\alpha\in\Omega^{0,1}(\End(V))$ are the bundle moduli, and the Beltrami differential $\mu\in\Omega^{0,1}(T^{1,0})$ gives the complex structure moduli. There is also an additional $B$-field component $b^{0,2}$ and an axio-dilaton field $c^{5,0}$. 

The fields $y^{0,3}$, $c^{5,2}$ and $b^{0,4}$ which are their anti-fields can then be seen as Lagrange multipliers and the part of the quadratic action featuring only these fields is
\begin{equation}
\begin{aligned}
\label{eq:ClassModAction}
    S &= \int_X \langle y^{0,1}, \bar{D} y^{0,3} \rangle \wedge \Omega 
   + \int_X \left(b^{0,2} \wedge \bar\der c^{5,2}+b^{0,4}\wedge\delb c^{5,0}\right)\\
   &\eqspace+ \int_X \left(y^{0,1} \inn \der b^{0,4}+y^{0,3} \inn \der b^{0,2}\right) \wedge \Omega\:.
\end{aligned}    
\end{equation}
The resulting equations of motion for $y^{0,1}$, $b^{0,2}$ and $c^{5,0}$, in analogy with the six-dimensional case \cite{Ashmore:2018ybe}, are the infinitesimal moduli equations for the ten-dimensional supersymmetric solution.

Writing the $y$-fields as $y=(x,\alpha,\mu)$, the gauge symmetry of \eqref{eq:ClassModAction} includes
\begin{equation}\label{eq:c_gauge}
    \delta\mu=\delb v\:,\qquad\delta c=\partial\left(v\inn\Omega\right)\:.
\end{equation}
By a Hodge decomposition, we can write $v$ as a total holomorphic derivative plus a divergence-free part,
\begin{equation}
    v^a=\nabla^aw + \tilde v^a\:,
\end{equation}
where $\nabla^a$ is the Chern connection and $\tilde v^a$ is divergence-free. Note that, as $\partial(v \inn \Omega) \propto (\nabla_a v^a)\Omega$, $\tilde v^a$ does not contribute to the gauge transformation of the $c$-fields. We can then use the total derivative $\nabla^a w$ to remove the $c$-fields from the action \eqref{eq:ClassModAction}. That is, the $c$-fields are pure gauge (up to harmonic components, which drop out of the action). If we also integrate out the $b$-fields by imposing their equation of motion, we find that the $\mu$-fields become divergence free,
\begin{equation}
    \nabla_a\mu^a=0\:.
\end{equation}
The remaining gauge symmetry for the $\mu$-fields comes solely from the divergence-free part of $v$, $\delta\mu=\delb\tilde v^a$. Such vectors $v$ generate the $\Omega$-preserving diffeomorphisms.

Following the above steps, the quadratic action for the geometric moduli simplifies to
\begin{equation} \label{eq:div-zero-action}
    S=\int_X \langle y^{0,1}, \bar{D} y^{0,3} \rangle \wedge \Omega\:,
\end{equation}
where the $\mu$-field contained in $y$ is now restricted to be divergence-free. Note that in the pairing on $\Omega^{0,\bullet}(Q)$
\begin{equation}
    \int_X \langle y, y' \rangle \wedge \Omega\:,
\end{equation}
the pairing of the $\der$-exact part of $x$ with the divergence part of $\mu'$ (i.e.~the total derivative $\LC^a w$ in the decomposition~\eqref{eq:c_gauge}) is zero. 
Thus, the $\der$-exact part of the $x$-fields drops out of the action~\eqref{eq:div-zero-action}. 
We may thus consider the fields $y$ to be elements of a quotient space
$\Pcomplex_\sim^{0,\bullet}$, which is defined by
\begin{equation} \label{P-def}
	\Pcomplex_\sim^{0,k} = \frac{\Pcomplex^{0,k}}{\Omega^{1,k}_{\der\text{-exact}}} 
	\quad \text{where} \quad
	\Pcomplex^{0,k} = \{ y \in \Omega^{0,k}(Q) \: : \: 
		\nabla_a \mu^a = 0 \}\subset \Omega^{0,\bullet}(Q)\:.
\end{equation}
This action then has a gauge symmetry given by shifting the $y$-fields by $\bar D$-exact terms in $\Pcomplex_\sim^{0,\bullet}$:
\begin{equation}
	\delta y = \bar{D} \epsilon \quad \text{for} \quad \epsilon \in \Pcomplex_\sim^{0,\bullet}\:.
\end{equation}

Clearly, we can perform the same gauge-fixing and integrating out procedure on the full field content of the theory from Section~\ref{sec:10d-theory} to obtain a classical master action
\begin{equation}
\label{eq:SheafTheory}
    S=\int_X \langle y, \bar{D} y \rangle \wedge \Omega\:,    
\end{equation}
where $y=y^{0,0}+\ldots+y^{0,5}$ is now a generic element of $\Pcomplex_\sim^{0,\bullet}$, with the odd fields bosonic and the even fields fermionic.\footnote{The field $y^{0,5}$ drops out of the quadratic theory, but does appear in the cubic theory below.} 
An alternative way to find the partition function~\eqref{eq:PartFunc} is then to note that, schematically, we have
\begin{equation}\label{eq:schematic_partition}
    \vert\bar D^{0,p}_{\Pcomplex} \vert = \frac{\vert\bar D^{0,p}\vert}{\vert\delb^{1,p}_{\del\text{-exact}}\vert\,\vert\delb^{5,p}_{\del\text{-exact}}\vert}\:,
\end{equation}
where $\vert\delb^{1,p}_{\del\text{-exact}}\vert$ is the determinant that one ``misses'' when we quotient $\Omega^{0,p}(\Lambda^{1,0})$ by $\del$-exact forms. Similarly, we have also removed the total holomorphic derivative part of $\Omega^{0,p}(T^{1,0})$ -- these are isomorphic to $\del$-exact $(5,p)$-forms. Note then that, ignoring harmonic forms, we can treat $\del$ as an isomorphism between $\Omega^{0,p}$ and $\Omega^{1,p}_{\del\text{-exact}}$. Furthermore, again ignoring harmonic forms, since $(5,p)$-forms are spanned by $\del$-exact forms, we can use the $\delb$-preserving isomorphism given by $\Omega$ to map these to $(0,p)$-forms. Hence we can identify
\begin{equation}
    \vert\delb^{1,p}_{\del\text{-exact}}\vert=\vert\delb^{0,p}\vert\:,\qquad\vert\delb^{5,p}_{\del\text{-exact}}\vert=\vert\delb^{5,p}\vert=\vert\delb^{0,p}\vert\:.
\end{equation}
The determinant in \eqref{eq:schematic_partition} then reduces to
\begin{equation}
    \vert\bar D^{0,p}_{\Pcomplex} \vert = \frac{\vert\bar D^{0,p}\vert}{\vert\delb^{0,p}\vert^2}\:,
\end{equation}
The corresponding product of determinants gives exactly the same result as \eqref{eq:PartFunc}.\footnote{This follows from writing $|\bar{\partial}^{0,p}|$ in terms of determinants of various $\Delta_{\bar\partial}^{0,q}$ operators, and then noting that $|\Delta_{\bar\partial}^{0,q}|=|\Delta_{\bar\partial}^{0,5-q}|$ implies $|\bar{\partial}^{0,0}|=|\bar{\partial}^{0,4}|$ and $|\bar{\partial}^{0,1}|=|\bar{\partial}^{0,3}|$.}\

\subsection{Generalising to a cubic theory}

The reason for following the steps to obtain \eqref{eq:SheafTheory} is that that theory has an obvious generalisation to a cubic Chern--Simons-like theory,
\begin{equation}
\label{eq:CubicSheafTheory}
    S=\int_X\left( \langle y, \bar{D} y \rangle-\tfrac{1}{3}\langle y,[y,y]\rangle\right) \wedge \Omega\:, 
\end{equation}
which is a ten-dimensional version of the action derived from the superpotential in \cite{Ashmore:2018ybe}.
The bracket $[\cdot,\cdot]$ on $\Pcomplex_\sim^{0,\bullet}$ is defined as
\begin{align}
    [y_1,y_2]^a&= \mu_{1}^{b}\wedge\partial_{b}\mu_{2}^{a}-(-1)^{|y_{1}||y_{2}|}\mu_{2}^{b}\wedge\partial_{b}\mu_{1}^{a} \:,\\
    [y_1,y_2]_{\alpha}    &=-\alpha_{1}\wedge\alpha_{2}+(-1)^{|y_{1}||y_{2}|}\alpha_{2}\wedge\alpha_{1}\nonumber\\
    &\eqspace+\mu_{1}^{b}\wedge\nabla_{b}\alpha_{2}-(-1)^{|y_{1}||y_{2}|}\mu_{2}^{b}\wedge\nabla_{b}\alpha_{1}\:,\\
    [y_1,y_2]_a&=2\,\mu_{1}^{b}\wedge\partial_{[b}x_{2,a]}-(-1)^{|y_{1}||y_{2}|}2\,\mu_{2}^{b}\wedge\partial_{[b}x_{1,a]}\nonumber\\
    &\eqspace+\tfrac{1}{2}\tr(\nabla_{a}\alpha_{1}\wedge\alpha_{2})-(-1)^{|y_{1}||y_{2}|}\tfrac{1}{2}\tr(\nabla_{a}\alpha_{2}\wedge\alpha_{1})\:,
\end{align}
where $y\in\Pcomplex_\sim^{0,p}$ of fermionic degree $f\in\{0,1\}$ has total degree $|y|=p+f$. In these expressions (and also~\eqref{eq:CubicSheafTheory} above), the wedge product should be understood as that compatible with the BV formalism, so that the bracket itself is graded commutative and satisfies $[y_1,y_2]=(-1)^{|y_1||y_2|}[y_2,y_1]$.\footnote{This wedge product satisfies $\alpha_1\wedge\alpha_2 = (-1)^{|\alpha_1||\alpha_2|}\alpha_2\wedge\alpha_1$. The sign here is different from what one would find by naively considering the boson/fermion number of the fields and taking the wedge product of the form components. The $|\alpha_1||\alpha_2|$-dependent sign can be thought of as coming from how one constructs a DGLA from a DGLA tensored with a differential graded commutative algebra. Alternatively, it is the same sign convention that one finds after promoting the fields to superfields by taking $\dd \bar{z}^a \mapsto \bar\theta^a$, where the $\bar\theta^a$ are odd coordinates on the supermanifold $T[1]X$.} 
This bracket satisfies the graded Jacobi identity and a Leibniz rule with respect to a version of the $\bar D$ operator similarly adapted to the BV formalism via $\bar{D} \ra (-1)^f \bar{D}$, where $f$ is the fermion number of the field acted on. The cubic action \eqref{eq:CubicSheafTheory} is then invariant under the gauge symmetry
\begin{equation} \label{eq:cubicGauge}
    \delta y=\bar D\lambda-[y,\lambda]\:,
\end{equation}
where $\lambda\in\Pcomplex_\sim^{0,\bullet}$.

With these definitions, the equation of motion of this cubic Chern--Simons-like theory is
\begin{equation} \label{eq:cubicEOM}
    \bar{D} y-\tfrac{1}{2}[y,y]=0\:,
\end{equation}
which can be thought of as a generalised Maurer--Cartan equation for the system.\footnote{This equation follows either from a variation of the action \eqref{eq:CubicSheafTheory}, or by noting the action is identical in form to that of the $SU(3)$ theory in \cite{Ashmore:2018ybe}, where a similar Maurer--Cartan equation was found.} In particular, restricting to form-degree two, we find
\begin{equation} \label{eq:BV-moduli}
    \bar{D} y^{0,1}-\tfrac{1}{2}[y^{0,1},y^{0,1}]-[y^{0,0},y^{0,2}]=0\:.
\end{equation}
Upon setting the fermionic modes $y^{0,0}$ and $y^{0,2}$ to zero, this is the Maurer--Cartan equation of the supersymmetric solution. This gives a Hull--Strominger-like system for $\mathcal{N}=1$ supergravity coupled to Yang--Mills. If one includes the tangent bundle connection as part of the gauge degrees of freedom, then \eqref{eq:BV-moduli} is a ten-dimensional version of the Hull--Strominger system including the spurious tangent bundle degrees of freedom~\cite{Ashmore:2018ybe}.

Note that the theory \eqref{eq:CubicSheafTheory} is not naturally associated to the $L_\infty$ algebra for the geometric moduli problem in the usual way. 
Rather, we have a supergravity extension of the holomorphic Chern--Simons action of~\cite{Baulieu:2010ch}, whose analogue of Equation~\eqref{eq:BV-moduli} becomes the holomorphic bundle condition when one sets the fermions to zero. 
As there, we consider the ``classical'' field to be the fermion $y^{0,2}$, with bosonic BRST ghosts $y^{0,1}$ (the geometric deformation) and fermionic ghosts-for-ghosts $y^{0,0}$, while the $y^{0,k}$ for $k \geq 3$ are identified as anti-fields. 
We have only linear terms in the anti-fields in~\eqref{eq:CubicSheafTheory} and the gauge algebra~\eqref{eq:cubicGauge} closes off-shell such that no higher order terms in anti-fields are required. 
One can check that 
\begin{equation}
    \omega(\delta y,\delta y') = \int_X \langle \delta y', \delta y \rangle \wedge \Omega\:,
\end{equation}
defines an odd symplectic form and that the action $S$ gives rise to the odd Hamiltonian vector field $Q$ with
\begin{equation}
    Q y =  \bar{D} y-\tfrac{1}{2}[y,y]\:.
\end{equation}
This gives the BRST operator for the symmetry~\eqref{eq:cubicGauge} and the 
classical master equation then follows from gauge invariance of the Maurer--Cartan equation~\eqref{eq:cubicEOM}. One can thus interpret the theory as a kind of BV theory, though we note that due to the definition~\eqref{P-def}, the fields are more complicated objects than one would like (c.f.~\cite{Costello:2015xsa, Costello:2021kiv} where the fields are defined to be elements of $\ker\del$). On top of this issue, the theory does not have a very conventional form, as there is a term involving only $y^{0,1} (y^{0,2})^2$. 
Here $y^{0,1}$ is a ghost field, so this is not part of an initial classical action, but it also does not come from the usual construction of a classical master action for a gauge algebra which closes off-shell. Therefore, exactly as for the system in~\cite{Baulieu:2010ch}, it is not clear from what classical theory this BV action arises. It is, however, a kind of cubic extension of our quadratic BV theory. Moreover, it can be connected to type I BCOV theory \cite{Costello:2015xsa, Costello:2019jsy, Costello:2021kiv}, through a non-local field redefinition, as we will now show.

\subsection{Connections to Costello--Li and Costello--Williams}
Coupling of gauge theory to gravitational degrees of freedom has appeared elsewhere in the literature. Of particular note is the type I topological string, or type I BCOV, of Costello--Li \cite{Costello:2015xsa, Costello:2019jsy} and Costello--Williams \cite{Costello:2021kiv}. This theory, usually studied on more symmetric torsion-free backgrounds (such as $\mathbb{C}^5$) with fluxes turned off, is related to the theory in the present paper, at least at the classical level, via a non-local field redefinition. We will now briefly explain this relation. 

For simplicity, and to make the connection clear, in this subsection we turn off background fluxes by setting $\bar D=\delb$ and assume our ten-dimensional spacetime $X$ is K\"ahler Calabi-Yau, so that K\"ahler identities also hold. We also ignore harmonic forms as in the above. Focusing on the gravitational part of the theory \eqref{eq:CubicSheafTheory}, which reads
\begin{equation}
    \label{eq:GravAction}
    S(x,\mu)=2\int_X\left(x_a\wedge \delb \mu^a-\tfrac{1}{2}\mu^a\wedge \mu^b\wedge (\del x)_{ab}\right)\wedge\Omega\:.
\end{equation}
we define a new field 
\begin{equation}
    \eta=\del x\:,
\end{equation}
where now $\eta\in\Omega^{2,\bullet}(X)$ is $\del$-closed global form. Note in particular that $\eta$ is invariant under shifts of $x$ by $\del$-exact forms. Fixing this shift symmetry by taking $x$ to be $\del^\dagger$-exact, we can write
\begin{equation}
    x=\del^{-1}\eta=\Delta_{\del}^{-1}\del^\dagger\eta\:,
\end{equation}
where $\Delta_{\del}^{-1}$ is the Green function. Plugging this back into \eqref{eq:GravAction}, we find
\begin{equation}
    \label{eq:GravAction2}
    S(x,\mu)=2\int_X\left((\del^{-1}\eta)_a \wedge  \delb \mu^a-\tfrac{1}{2}\eta_{ab}\wedge \mu^a\wedge \mu^b\right)\wedge\Omega\:.
\end{equation}
Note that we may view $\eta$ as a field valued in $\Omega^{0,\bullet}(\Lambda^3T^{1,0})$, where the antiholomorphic top-form $\bar\Omega$ is used to raise the indices. The field $\eta$ is then in the kernel of the divergence operator $\del$ as defined in \cite{Costello:2015xsa, Costello:2019jsy, Costello:2021kiv}. Upon integrating by parts and using the K\"ahler identities, the action \eqref{eq:GravAction2} then becomes precisely the gravitational action, or closed string action, of Costello--Li--Williams (modulo some minus signs and irrelevant numerical factors which can be absorbed by field redefinitions). 

The coupling of the theory to the gauge sector is then exactly the same as in Costello--Li--Williams, and the classical BV theories agree, modulo this non-local field redefinition. Given the non-local nature of the field redefinition, one might wonder if they also agree at the quantum level, or indeed if they are related by some $L_\infty$ quasi-isomorphism. We expect this to be the case, given that Costello--Li--Williams also conjecture their theory to be twisted type I supergravity, or twisted supergravity coupled to Yang--Mills. In particular, as we shall see in the next section, our theory has similar gauge and gravitational anomalies as the ones found in \cite{Costello:2015xsa, Costello:2019jsy, Costello:2021kiv}.

\section{Anomalies}\label{sec:anomalies}

As shown in Equation \eqref{eq:PartFunc}, the absolute value of the one-loop partition function of \eqref{eq:BVaction} is
\begin{equation}
\label{eq:OneLoop2}
    \vert Z\vert^2=\frac{I_{\bar D}(Q)}{I(\Lambda^{0,0})^2}, 
\end{equation}
where $Q=\Lambda^{1,0} \oplus \End(V) \oplus T^{1,0}$, $\End(V)$ may or may not include extra $\End(T)$-valued spurious degrees of freedom, and $\bar D$ is a holomorphic differential with off-diagonal flux terms (see \eqref{eq:barD1}). Note that the theory \eqref{eq:BVaction} may be defined on any classical solution to the ten-dimensional Hull--Strominger system, torsional or not, and the corresponding one-loop partition function \eqref{eq:OneLoop2} may be defined, and possibly even computed explicitly.

When studying the quantum theory that leads to \eqref{eq:OneLoop2}, we should also consider its anomalies. Anomalies are associated to an anomalous phase transformation of the partition function, and so do not appear in the absolute value $|Z|$. Anomalies can occur for global symmetries, local (gauge) symmetries, or changes of the background fields defining the underlying geometry. Of these, only anomalies in local symmetries are ``fatal'', indicating that the theory is inconsistent. We compute these anomalies in the coming subsections, and then suggest a number of ways to cancel them.

It turns out that these anomalies change depending on whether the spurious $\End(T)$-valued modes are included or not. As these modes are non-physical, we do not include them in what follows. As noted above, this forces classical solutions to the ten-dimensional heterotic supersymmetry equations on compact geometries to be K\"ahler and Calabi--Yau, where the operator on $Q$ is diagonal, $\bar D=\delb$, and the fluxes drop out at the classic level where $\hbar=0$. At the quantum (one-loop) level, we will see how imposing anomaly cancellation recovers the upper-triangular extension structure of the operator on $Q$, though the extension classes are no longer tensorial. 

\subsection{Anomaly polynomials}
As noted above, the relevant local anomalies are related to an anomalous transformation of the phase of the partition function. Indeed, the partition function $Z$ should be interpreted as a section of a certain holomorphic determinant line bundle over the complex configuration space $\cal M$ of geometric structures on $X$. If the Chern connection on this bundle has non-zero curvature, ${\cal F}_{\rm Det}\neq0$, then the phase of $Z$ is not fully specified by the gauge-invariant data of a point in the configuration space. This implies there is an anomaly.

Adapting the results of \cite{bismut1988analytic1,bismut1988analytic2,bismut1988analytic3,Bittleston:2022nfr} to our setting, the curvature ${\cal F}_{\text{Det}}\neq0$ reads
\begin{equation}
\label{eq:AnomPol}
    {\cal F}_{\rm Det}=\frac{1}{2}\left[\int_X\op{td}(X)\wedge\op{ch}(Q)\right]_{(1,1)}-\left[\int_X\op{td}(X)\right]_{(1,1)}\coloneqq\left[-\int_X{\cal P}\right]_{(1,1)}\:,
\end{equation}
where the characteristic polynomials $\op{ch}(Q)$ and $\op{td}(X)$ are constructed using curvatures on the ``universal geometry''~\cite{Candelas:2018lib, Garcia-Fernandez:2018emx, McOrist:2019mxh, McOrist:2024glz, Tellez-Dominguez:2023wwr, dominguez2025moduli}, given by the total space of a family of heterotic vacua fibered over the parameter space, $X \to \mathcal{M}$. The integration over $X$ should be read as an integration over the fibres, resulting in a $(1,1)$-form on $\cal M$. Since $X$ is ten-dimensional, only the $(6,6)$-component (as a form on the total space) of the curvature polynomial $\cal P$ contributes to ${\cal F}_{\rm Det}$.

Given that $Q = \Lambda^{1,0} \oplus \End(V) \oplus T^{1,0}$, the total curvature polynomial is
\begin{align}
{\cal P}_{(6,6)} & =\frac{\dim(G)-496}{252}\ch_{6}(X)+\frac{\dim(G)-64}{10368}\ch_{2}(X)^{3}\nonumber \\
 & \eqspace+\frac{\dim(G)+224}{1440}\ch_{2}(X)\ch_{4}(X)-\tfrac{1}{288}\ch_{2}(X)^{2}\ch_{2}(\End(V))\\
 & \eqspace-\tfrac{1}{120}\ch_{4}(X)\ch_{2}(\End(V))+\tfrac{1}{12}\ch_{2}(X)\ch_{4}(\End(V))-\ch_{6}(\End(V))\:.\nonumber 
\end{align}
To cancel the resulting anomaly using a Green--Schwarz-like mechanism, we must factorise the curvature polynomial. This cannot be done in general. We first restrict to bundles $V$ whose structure groups have dimension 496, $\dim G = 496$, so that the $\ch_6(X)$ term drops out.
Doing this, the polynomial simplifies to
\begin{align}
    {\cal P}_{(6,6)}&=\tfrac{1}{24}\ch_2(X)^3 + \tfrac{1}{2} \ch_2(X) \ch_4(X) - \tfrac{1}{288} \ch_2(X)^2 \ch_2(\End(V))\\ 
    & \eqspace - \tfrac{1}{120} \ch_4(X) \ch_2(\End(V))+ \tfrac{1}{12}\ch_2(X) \ch_4(\End(V))  - \ch_6(\End(V))\:.\nonumber
\end{align}
This polynomial can be factorised for the gauge groups $SO(32)$ and $E_8\times E_8$, which both have $\dim G = 496$, just as in heterotic string theory. 

For $SO(32)$, the polynomial simplifies to
\begin{align}
    {\cal P}_{(6,6)}&=\tfrac{1}{24} \ch_2(X)^3 + \tfrac{1}{2} \ch_2(X) \ch_4(X) - \tfrac{5}{48} \ch_2(X)^2 \ch_2(V) - \tfrac{1}{4} \ch_4(X) \ch_2(V)\notag\\
    &\eqspace + \tfrac{1}{24}\ch_2(X) \ch_2(V)^2  + 2 \ch_2(X) \ch_4(V) - \ch_2(V)\ch_4(V)\:,
\end{align}
which then factorises as
\begin{align}\label{eq:SO32_anomaly}
    {\cal P}_{(6,6)}&=\tfrac{1}{24} \bigl(\ch_2(X)-\tfrac12 \ch_2(V) \bigr)\bigl(\ch_2(X)^2 + 12 \ch_4(X)\notag\\
    &\eqspace- 2\ch_2(X)\ch_2(V)  + 48 \ch_4(V)\bigr)\notag\\
    &=\tfrac{1}{24} \bigl(\ch_2(X)-\tfrac12 \ch_2(V) \bigr)\wedge{\cal P}^{SO(32)}_{(4,4)}\:,
\end{align}
where we have defined the degree-$(4,4)$ curvature polynomial ${\cal P}^{SO(32)}_{(4,4)}$ along the way. 

Choosing $E_8 \times E_8$ instead, we find
\begin{align}
    {\cal P}_{(6,6)}&=\tfrac{1}{24} \ch_2(X)^3 + \tfrac12 \ch_2(X) \ch_4(X) - \tfrac{5}{48} \ch_2(X)^2 \ch_2(V_1)  - \tfrac14 \ch_4(X)\ch_2(V_1) \notag\\
    &\eqspace+ \tfrac{1}{8}\ch_2(X) \ch_2(V_1)^2  - \tfrac{1}{24}\ch_2(V_1)^3 - \tfrac{5}{48} \ch_2(X)^2 \ch_2(V_2)  \notag\\
    &\eqspace- \tfrac14 \ch_4(X) \ch_2(V_2) + \tfrac{1}{8}\ch_2(X) \ch_2(V_2)^2 - \tfrac{1}{24}\ch_2(V_2)^3\:,
\end{align}
where $V_1$ is the bundle corresponding to the first $E_8$ factor, and $V_2$ is the bundle corresponding to the second $E_8$ factor. This polynomial factorises as
\begin{align}\label{eq:E8_anomaly}
   {\cal P}_{(6,6)}&=\tfrac{1}{48}\bigl(\ch_2(X)-\tfrac{1}{2}\ch_2(V)\bigr) \bigl(2 \ch_2(X)^2 - 24 \ch_4(X) - 4 \ch_2(X) \ch_2(V_1)  \notag\\ 
   &\eqspace + 4\ch_2(V_1)^2+4 \ch_2(X) \ch_2(V_2) - 4 \ch_2(V_1) \ch_2(V_2) + 4\ch_2(V_2)^2 \bigr)\notag\\
   &=\tfrac{1}{48}\bigl(\ch_2(X)-\tfrac{1}{2}\ch_2(V)\bigr)\wedge{\cal P}^{E_8\times E_8}_{(4,4)}\:,
\end{align}
where this again serves as a definition of the degree-$(4,4)$ curvature polynomial ${\cal P}^{E_8\times E_8}_{(4,4)}$. 
Note that these are the same anomaly polynomials that appear in heterotic supergravity (see, for example, \cite{Bilal:2008qx}), supporting the conjecture that the holomorphic theory we are considering is indeed twisted ten-dimensional supergravity coupled to Yang--Mills.

\subsection{Cancelling anomalies}

Focusing on $SO(32)$ and $E_8\times E_8$, we now attempt to cancel the anomalies using a Green--Schwarz-type mechanism. By the descent procedure, the anomaly polynomials lead to anomalous gauge transformations of the partition function.\footnote{See \cite{Bilal:2008qx} for an introduction to anomalies in field theory and gravity, and the descent procedure.} Specifically, we can write the anomaly polynomial ${\cal P}_{(6,6)}={\cal P}_{12}$ as
\begin{equation}
    {\cal P}_{12}=\dd{\cal Q}_{11}\:,
\end{equation}
for some Chern--Simons-like eleven-form ${\cal Q}_{11}$. Furthermore, ${\cal Q}_{11}$ changes under gauge transformations by a exact form
\begin{equation}
    \delta{\cal Q}_{11}=\dd{\cal P}_{10}(\gamma,\epsilon)\:,
\end{equation}
where $\epsilon$ is the infinitesimal gauge transformation of the $G$-valued gauge field, while $\gamma^a{}_b=\nabla_b v^a$ is a change of frame corresponding to an infinitesimal holomorphic diffeomorphism $v^a$. The anomalous transformation of the partition function is then given as
\begin{equation}
    \delta\log{Z}=-\ii\int_X{\cal P}_{10}(\gamma,\epsilon)\:.
\end{equation}
Of course, there are choices involved in ${\cal Q}_{11}$ and ${\cal P}_{10}(\gamma,\epsilon)$. These choices are related to which phase convention we pick for the partition function. 

The descent procedure leading to the usual Green--Schwarz anomalous transformation then gives\footnote{Recall that the generators of $SU(N)$, $SO(32)$ and $E_8$ are conventionally normalised so that the trace of their square is $1/2$, $1$ and $30$ in the $\rep{N}$, $\rep{32}$ and $\rep{248}$ representations respectively. For $E_8$, a factor of $1/30$ is then included in the definition of the trace and then referred to as the trace in the ``fundamental'' representation. The $SU(5)$ generators in the $\rep{5}$ representation that give $\ch_2(X)$ can be embedded as $\rep{5}\oplus\repb{5}$ in the $\rep{10}$ of $SO(10)$, where one should include a factor of $1/2$ so that $\tr_{\rep{5}}(T^2) = \tfrac12\tr_{\rep{10}}(T^2)$. The result of this is that the relative factor of $1/2$ in both \eqref{eq:SO32_anomaly} and \eqref{eq:E8_anomaly} drops out.}
\begin{equation}
\label{eq:AnomalousPhase}
    \delta\log{Z}=\ii\beta'\int_X\bigl(\tr(R\gamma)-\tr(F\epsilon)\bigr)\wedge{\cal P}_{(4,4)}\:,
\end{equation}
where the various prefactors have been collected into a constant which we have called $\beta'$, anticipating its connection with the $\alpha'$ parameter of heterotic supergravity. Note that the $\beta'$ defined above includes an implicit factor of $\hbar$, indicating the quantum nature of the anomaly. 

There are several ways to cancel the anomaly. We first consider using non-local counter-terms. This is perhaps the most straightforward method, but the least satisfactory, as it renders the full theory non-local. We then use a Green--Schwarz-like mechanism, where a choice of harmonic gauge again forces a mild non-locality on us. Next, we use local counter-terms which come from correcting the hermitian Yang--Mills equation to a deformed instanton equation~\cite{Marino:1999af, Leung:2000zv}. This type of deformation is expected at one-loop order, so it is not too surprising that it naturally appears. Finally, we consider non-global cancellation, where the counter-terms are local, but need not be global in general. This hints towards interesting new geometric structures, which we hope to explore in future works. We give a summary and comparison of the four different methods of cancelling the anomaly in Section \ref{sec:Comp}.

\subsection{Non-local cancellation}

The $T^{1,0}$-valued part of $y^{0,1}$, which we have previously written as $\mu^a$, can be interpreted as a complex structure deformation. Under a gauge transformation by $\gamma^a{}_b=\nabla_b v^a$, $\mu^a$ transforms as
\begin{equation}
    \delta\mu^a=\delb v^a\:.
\end{equation}
The $\gamma$-dependent term of the anomalous transformation \eqref{eq:AnomalousPhase} can then be cancelled by adding the counter-term
\begin{equation}
    S_{\rm counter}(\mu)=-\beta'\int_X {R^a}_b\nabla_a\left(\Delta^{-1}_{\delb}\delb^\dagger\mu^b\right)\wedge{\cal P}_{(4,4)}
\end{equation}
to the classical action. Here, $\Delta^{-1}_{\delb}$ is the Green function of the $\delb$-Laplacian, so the counter-term is explicitly non-local on $X$. Moreover, since the Laplace operator depends on a choice of background metric, one expects the resulting theory to have a metric anomaly, i.e.~depend on the choice of metric.

Similarly, the $\End(V)$-valued part of $y^{1,0}$, written as $\alpha$, may be interpreted as a holomorphic deformation of the gauge bundle. This transforms as
\begin{equation}
    \delta\alpha=\delb_A\epsilon\:.
\end{equation}
The $\epsilon$-dependent term of \eqref{eq:AnomalousPhase} can then be cancelled by the non-local counter-term
\begin{equation}
    S_{\rm counter}(\alpha)=\beta'\int_X\tr\bigl(F\Delta^{-1}_{\delb_A}\delb_A^\dagger\alpha\bigr)\wedge{\cal P}_{(4,4)}\:,
\end{equation}
where $\Delta^{-1}_{\delb_A}$ is the Green function of the $\delb_A$-Laplacian. Note that this counter-term will not absorb an anomalous gauge transformations when $\epsilon$ itself is holomorphic, i.e.~$\delb_A\epsilon=0$. If, however, the bundle $V$ is stable, no such sections exist. Since stability is necessary for the undeformed background to be supersymmetric, we can assume that all anomalous gauge transformations can be cancelled using the above counter-term.\footnote{Alternatively, since $\delb_A\epsilon=0$ implies that $\epsilon$ is a harmonic section, and we have dropped harmonic forms in our quantum computations, we should also drop such holomorphic sections.}

\subsection{Anomaly cancellation with the usual Green--Schwarz mechanism}
Though straightforward, the approach used in the previous subsection is somewhat unsatisfactory due to the non-local nature of the counter-term and an implicit dependence on the background metric on $X$ in the Green functions and the adjoint differential operators. It would be preferable to cancel the anomalies using only local counter-terms while retaining as much of the topological nature of the theory as possible. In this section, we will cancel the anomaly using a version of the more familiar Green--Schwarz mechanism \cite{Green:1984sg}. We will find that in order to have an exactly gauge invariant one-loop theory, we need to adjust the kinetic operator $\delb$ to a new operator $\DD$, which squares to zero precisely when the heterotic Bianchi identity is satisfied. This adjustment however introduces new gauge transformations under which the Green--Schwarz counter-terms are not invariant. This is resolved by the addition of further counter-terms.  

In analogy with the Green--Schwarz mechanism, we add a correction to the gauge transformation \eqref{eq:gauge_y} for the hermitian degrees of freedom:
\begin{equation}
\label{eq:AnomalousGaugex}
    \delta x= \beta'\bigl(\tr(F\epsilon)-\tr(R\gamma)\bigr)\:,
\end{equation}
and add a counter-term of the form
\begin{equation}
\label{eq:counter-x}
    S_\text{counter}(x)=\int_X x\wedge{\cal P}_{(4,4)}\:.
\end{equation}
Upon including the gauge transformation \eqref{eq:AnomalousGaugex}, the original classical action with a diagonal $\delb$-operator is no longer gauge invariant. However, the gauge transformation \eqref{eq:AnomalousGaugex} is of one-loop order ${\cal O}(\beta')$, which renders the other fields classical under such a transformation, and therefore $\bar\der$-closed modulo higher loop effects. That is, since the gauge transformation is itself order $\beta'$, when computing the gauge transformation of the action to one-loop order ${\cal O}(\beta')$, the terms that multiply this transformation include only the classical ${\cal O}(1)$ parts of the other fields. These classical parts are $\delb$-closed, and so the classical action is hence still gauge invariant modulo higher loop effects.

Despite this, we would still like to find an action that is exactly gauge invariant, not just modulo higher-loop effects. Indeed, as we shall see, this will lead to interesting new geometric structures. To do this, we first change the diagonal kinetic operator $\delb$ in the classical action on the bundle $Q$ to an operator $\DD$ given by
\begin{equation}
\label{eq:barD-SE}
    \DD = 
\begin{pmatrix}
\delb & \,\beta'\mathcal{F}^*\, &  \beta'R \cdot \nabla\\
0 & \delb_\A & \mathcal{F}\\
0 & 0 & \delb 
\end{pmatrix}\:,
\end{equation}
by adding appropriate extra counter-terms.\footnote{Strictly speaking, the added terms adjusting $\delb\rightarrow\DD$ are not counter-terms in the usual sense. An alternative point of view is to use the operator \eqref{eq:barD-SE} in the classical theory. Though not a connection, $\DD$ does define an extension structure on the sum of bundles given by $Q$. The operator $\DD$ has the same index formula as the diagonal operator $\delb$ on $Q$ \cite{Chisamanga:2024xbm, deLazari:2024zkg, Ibarra:2024ntg}, and the characteristic classes making up the anomaly polynomial should hence still decompose into the individual components, so that the anomaly polynomials remain the same. We will investigate this in future work.} This is a different operator from $\bar{D}$ used above, and will be used to define the kinetic operators in this section.

The off-diagonal components of $\DD$ act on the field $y=(x,\alpha,\mu) \in \Omega^{0,\bullet}(Q)$ as
\begin{align}
    \mathcal{F}(\mu)&=F_{a}\wedge\mu^a\:,\\
    \left(\mathcal{F}^*(\alpha)\right)_a&=\tr(F_{a}\wedge\alpha)\:,\\
    \left(R\cdot\nabla\mu\right)_a&=-R_a{}^b{}_c\wedge\nabla_b\mu^c\:,
\end{align}
where $F_a=F_{a\bar b}\dd z^{\bar b}$ is the gauge field strength, and $R_a{}^b{}_c=R_{a\bar b}{}^b{}_c\,\dd z^{\bar b}$ is the $\End(T^{1,0})$-valued curvature two-form of the Levi--Civita connection of the background K\"ahler metric. This kind of operator  appears when studying moduli problem of the Hull--Strominger system without the extra spurious degrees of freedom~\cite{McOrist:2021dnd}; its mathematical properties have been studied more recently~\cite{McOrist:2024zdz, Chisamanga:2024xbm, Ibarra:2024ntg}. 

The new master action
\begin{equation}
\label{eq:BVaction2}
    S=\int_X\langle y,\DD y\rangle\wedge\Omega+\int_X b\wedge \bar\partial c + \int_X (y \inn \der b) \wedge \Omega
\end{equation}
is then invariant under the gauge transformation $\delta y=\DD \gamma$ provided $\DD^2=0$. With $\DD$ defined by \eqref{eq:barD-SE}, nilpotency requires~\cite{McOrist:2021dnd}
\begin{equation}
\label{eq:exactBI}
    \DD^2=0\quad\Rightarrow\quad\tr\left(F\wedge F\right)=\tr\left(R\wedge R\right)\:.
\end{equation}
Crucially, this is an equation on four-forms rather than on cohomology classes. This is too strong for our purposes -- in general \eqref{eq:exactBI} should hold only in cohomology for the Hull--Strominger as it is related to $\dd H$~\cite{Strominger:1986uh, Hull:1986kz, Garcia-Fernandez:2016azr}.\footnote{There are non-generic examples such as the standard embedding where it does hold as an equation on four-forms as $\dd H = 0$~\cite{Candelas:1985en}.} We can relax this by further adjusting the differential $\DD$. 

To see how to do this, we start from our goal by assuming that \eqref{eq:exactBI} is satisfied in cohomology, which implies that
\begin{equation}
\label{eq:BI}
    \tr(F\wedge F)-\tr(R\wedge R)=i\partial\bar\partial\tau\:,
\end{equation}
where $\tau$ is a real $(1,1)$-form. This equation fixes $\tau$ in terms of the background curvatures, up to $\del\delb$-closed terms. We then identify $\beta'$ with the usual string tension $\alpha'$, and define a corrected hermitian form 
\begin{equation}
\label{eq:Corr-Omega}
    \omega=\omega_0+\frac{\beta'}{4}\tau\:,
\end{equation}
where $\omega_0$ is the K\"ahler form of the background Calabi--Yau. Our starting Equation \eqref{eq:BI} then becomes the usual heterotic Bianchi identity,
\begin{equation}
\label{eq:BI2}
    \ii\partial\bar\partial\omega=\frac{\beta'}{4}\bigl(\tr(F\wedge F)-\tr(R\wedge R)\bigr)\:.
\end{equation}
We then correct the $\DD$ operator to
\begin{equation}
\label{eq:barD}
    \DD=
    \begin{pmatrix}
    \delb & \,\beta'\mathcal{F}^*\;\; &  \mathcal{T}+\beta'R \cdot \nabla\\
    0 & \delb_\A & \mathcal{F}\\
    0 & 0 & \delb 
    \end{pmatrix}\:,
\end{equation}
where the operator $\mathcal{T}$ acts as
\begin{equation}
    \left(\mathcal{T}(\mu)\right)_a=-4\ii\,\partial_{[a}\omega_{b]}\wedge\mu^b=-\ii\beta'\,\partial_{[a}\tau_{b]}\wedge\mu^b\:,
\end{equation}
with $\omega_b=\omega_{b\bar c}\dd z^{\bar c}$ and $\tau_b=\tau_{b\bar c}\dd z^{\bar c}$. With this redefinition, nilpotency of $\DD$ implies only the weaker cohomological condition \eqref{eq:BI}, and the master action \eqref{eq:BVaction2} becomes invariant under $\DD$-exact gauge transformations.

The new $\DD$-exact gauge transformations do not however leave the full one-loop theory invariant (the master action plus counter-terms). Specifically, the counter-term \eqref{eq:counter-x} will pick up an additional transformation
\begin{equation}
    \delta S_{\text{counter}}(x)
    =2\ii\int_X (v\inn\partial\omega)\wedge{\cal P}_{(4,4)}=2\ii\int_X {\cal L}_v \omega\wedge{\cal P}_{(4,4)}\:, 
\end{equation}
which then needs to be cancelled by additional counter-terms. Here ${\cal L}_v$ is the holomorphic Lie derivative, defined by
\begin{equation}
    {\cal L}_v\beta=v\inn\partial\beta+\partial(v\inn\beta)
\end{equation}
for any form $\beta$.

To see how to cancel this transformation, let us first decompose ${\cal P}_{(4,4)}$ with respect to the hermitian $(1,1)$-form \eqref{eq:Corr-Omega} as
\begin{equation}
    {\cal P}_{(4,4)}=\tfrac{1}{3!}\omega^3\wedge{\cal P}^\text{p}_{(1,1)}+\tfrac{1}{4!}\omega^4\,{\cal P}_0\:,
\end{equation}
which also implies that
\begin{equation}
    \star{\cal P}_{(4,4)}=-{\cal P}^\text{p}_{(1,1)}+\omega\,{\cal P}_0\:,
\end{equation}
where $\star$ is the Hodge dual associated to the metric defined by $\omega$ and the background complex structure on $X$. Here, the $(1,1)$-form ${\cal P}^\text{p}_{(1,1)}$ is primitive, i.e.~$\omega\inn{\cal P}_{(1,1)}=0$. The variation of the counter-term can then be written as
\begin{align}
    \delta S_\text{counter}(x)&=2\ii\int_X {\cal L}_v\left(\tfrac{1}{4!}\omega^4\right)\wedge\left({\cal P}^p_{(1,1)}+\tfrac{1}{4}\omega\,{\cal P}_0\right)\notag\\
    &=2\ii\int_X{\cal L}_v\left(\tfrac{1}{4!}\omega^4\right)\wedge\left(-{\star{\cal P}}_{(4,4)}+\tfrac{5}{4}\omega\,{\cal P}_0\right)\:.\label{eq:counter_1}
\end{align}

To further develop the above expression, we need to consider how supersymmetry constrains the undeformed background geometry. In particular, the background should solve the hermitian Yang--Mills equation 
\begin{equation}
    \omega^4\wedge F=0\:,
\end{equation}
with a hermitian form $\omega$ that is conformally balanced,
\begin{equation}
    \dd(\ee^{-2\phi}\omega^4)=0\:,
\end{equation}
where $\phi$ is the dilaton. For even-dimensional heterotic solutions where a zeroth order K\"ahler Calabi--Yau background exists (or more generally a torsion-free zeroth-order background), we may without loss of generality assume that the dilaton is constant. Indeed, it was shown in \cite{Anguelova:2010ed, McOrist:2025zwf} that for six-dimensional heterotic solutions, the dilaton may be set to a constant modulo ${\cal O}(\alpha'^3)$ corrections. Their argument can be adapted to other dimensions. 
In even dimensions, this implies that $\omega$ is co-closed
\begin{equation}    
    \dd\star\omega=\dd\left(\tfrac{1}{4!}\omega^4\right)=0\:,
\end{equation}
which is equivalent to $\dd^\dagger\omega=0$ as $\dd^\dagger=-\star\dd\star$.

Using that $\omega$ is co-closed, it follows that the holomorphic Lie derivative term in \eqref{eq:counter_1} can be expanded as
\begin{equation}
    {\cal L}_v\left(\tfrac{1}{4!}\omega^4\right)=\partial\left( v\inn\tfrac{1}{4!}\omega^4\right)=\partial\left((v\inn\omega)\wedge\tfrac{1}{3!}\omega^3\right)\:.
\end{equation}
Next, note that changing the representatives of the characteristic classes appearing in the anomaly polynomial $\cal P$ in \eqref{eq:AnomPol} only serves to change the curvature ${\cal F}_{\rm Det}$ by irrelevant exact terms. Indeed, as we discuss further below, these terms only serve to change the phase convention that we pick for the partition function. Indeed, it is only the $\bar\partial$-harmonic part of ${\cal P}_{(4,4)}$ is relevant for the anomaly (where harmonic is with respect to the non-K\"ahler metric given by $\omega$).
If we then pick a phase convention for $Z$ where ${\cal P}_{(4,4)}$ appearing in \eqref{eq:AnomalousPhase} is $\bar\partial$-harmonic, we find
\begin{align}
    \delta S_\text{counter}(x)&=\tfrac{5}{2}\ii\int_X\partial\left((v\inn\omega)\wedge\tfrac{1}{3!}\omega^3\right)\wedge\omega\,{\cal P}_0=10\ii\int_X\partial(v\inn\omega)\wedge\tfrac{1}{4!}\omega^4\,{\cal P}_0\notag\\
    &=10\ii\int_X {\cal L}_v\left(\tfrac{1}{5!}\omega^5\right)\,{\cal P}_0=10\ii\int_X \tfrac{1}{\vert\Omega\vert^2}{\cal L}_v\Omega\wedge\bar\Omega\,{\cal P}_0\:,
\end{align}
where in the first equality the term involving $\star{\cal P}_{(4,4)}$ drops out by integration by parts, as 
\begin{equation}
    \delb^\dagger{\cal P}_{(4,4)}=-\star\del\star{\cal P}_{(4,4)}=0\:,
\end{equation}
which implies $\del\star{\cal P}_{(4,4)}=0$. We have again also used that $\omega$ is co-closed, and the $SU(5)$ structure relation
\begin{equation}
    \tfrac{\ii}{\vert\Omega\vert^2}\Omega\wedge\bar\Omega=\tfrac{1}{5!}\omega^5\:.
\end{equation}
Note that the norm of $\Omega$, which for supersymmetric background solutions is given by
\begin{equation}
\vert\Omega\vert=\ee^{-2\phi}\:,    
\end{equation}
is constant when the dilaton is constant, which is true when $\omega$ is co-closed. 

Looking back to \eqref{eq:c_gauge}, the gauge transformation of the axio-dilaton field fluctuation $c^{5,0}$ of the classical action \eqref{eq:BVaction2} is
\begin{equation}
    \delta c^{5,0}=\partial(v\inn\Omega) \equiv {\cal L}_v \Omega\:.
\end{equation}
So we can cancel the transformation of $S_{\rm counter}(x)$ by adding the counter-term
\begin{equation}
    S_{\text{counter}}(c^{5,0})=-10\ii\int_X \tfrac{1}{\vert\Omega\vert^2}c^{5,0}\wedge\bar\Omega\,{\cal P}_0\:.
\end{equation}
With this addition, the full one-loop theory will be gauge invariant, where the one-loop action reads
\begin{align}
    S_{\text{one-loop}}&=S+S_{\text{counter}}(x)+S_{\text{counter}}(c^{5,0})\:,
\end{align}
where $S$ is given by \eqref{eq:BVaction2}.

It should however be noted that we had to pick a non-local convention for the phase of the partition function, in that we chose ${\cal P}_{(4,4)}$ to be {\it harmonic}. This involves doing a Hodge decomposition of ${\cal P}_{(4,4)}$ and then projecting onto its harmonic part, which is an intrinsically non-local operation. In the phase convention where we use hermitian Yang--Mills curvatures, it seems difficult to cancel the anomalies exactly without making such a non-local choice. In the next section we will remedy this, by considering {\it deformed instantons} instead.

\subsection{Local anomaly cancellation with deformed instantons}\label{sec:deformed_instantons}

Let us recall the anomalous transformation of the partition function in \eqref{eq:AnomalousPhase}:
\begin{equation}
    \delta\log{Z}=\ii\beta'\int_X\bigl(\tr(R\gamma)-\tr(F\epsilon)\bigr)\wedge{\cal P}_{(4,4)}\:.
\end{equation}
As in the previous subsection, the curvatures here are the hermitian Yang--Mills curvatures of the holomorphic tangent bundle (i.e.~the Ricci-flat curvature of the Calabi--Yau metric) and the holomorphic vector bundle. The hermitian Yang--Mills equation is, however, expected to receive corrections at one-loop order, changing it to a {\it deformed instanton} equation \cite{Marino:1999af, Leung:2000zv}.\footnote{See also \cite{deBoer:2006bp}, where the one-loop effective action of a seven-dimensional Chern--Simons theory on a $G_2$ manifold corrects the equation of motion to a deformed instanton.} As we shall see, a convenient choice of deformed instanton equations to consider is
\begin{align}
    \label{eq:DefInst1}
    \tfrac{1}{4!}\omega^4\wedge\tilde F&={\cal P}_{(4,4)}(\tilde F,\tilde R)\wedge \tilde F\:, %
    \\
    \label{eq:DefInst2}
    \tfrac{1}{4!}\omega^4\wedge\tilde R&={\cal P}_{(4,4)}(\tilde F,\tilde R)\wedge \tilde R%
    \:,
\end{align}
where $\tilde F$ and $\tilde R$ are the curvatures of deformed hermitian metrics $\tilde h$ and $\tilde g$ on the vector bundle and tangent bundle respectively. It would be interesting to understand whether there is a physical motivation for considering the specific equations above. The reader may also wonder if these deformed instanton equations have solutions -- we will comment on this below. 

Modulo higher-curvature corrections,\footnote{Note that the deformed instanton equations \eqref{eq:DefInst1}--\eqref{eq:DefInst2} correct the instanton equation at fifth order in the curvature. This is the same order as quantum corrections, such as the anomalous phase, hence we can include it as an ${\cal O}(\beta')$ quantum correction of the hermitian Yang--Mills constraint.} the anomalous transformation of the partition function can be written as\footnote{Alternatively, this can be viewed as a phase convention for the partition function where we use the curvatures $\tilde R$ and $\tilde F$ instead.}
\begin{align}
    \delta\log{Z}&=\ii\beta'\int_X\bigl(\tr(\tilde R\gamma)-\tr(\tilde F\epsilon)\bigr)\wedge{\cal P}_{(4,4)}(\tilde F,\tilde R)\notag\\
    &=\ii\beta'\int_X\bigl(\tr(\tilde R\gamma)-\tr(\tilde F\epsilon)\bigr)\wedge\tfrac{1}{4!}\omega^4
    \:,
    \label{eq:AnomPhase2}
\end{align}
where we have used the deformed instanton equation to reach the final line. With this anomalous phase, we change the counter-term \eqref{eq:counter-x} to read
\begin{equation}
\label{eq:CounterX2}
    S^2_{\rm counter}(x)=\frac{1}{4!}\int_X x\wedge\omega^4\:.
\end{equation}
Note that, again assuming that $\omega$ is co-closed, this counter-term is invariant under $\partial$- and $\bar\partial$-exact shifts of $x$.

It is also interesting to note that the counter-term \eqref{eq:CounterX2} is similar in form to the correction needed to cancel the metric anomaly of the one-loop partition function of the type IIB generalised Hitchin functional \cite{Pestun:2005rp}. Indeed, it corresponds to a variation of the K\"ahler potential for the hermitian degrees of freedom:
\begin{equation}
    \ee^{-K[\omega]}=\frac{1}{5!}\int_X\omega^5\:.
\end{equation}
We hope to investigate this connection further in future works when we study the possible anomalous metric dependence of the partition function.

We also need to promote the curvatures in the $\DD$ operator \eqref{eq:barD} to $\tilde F$ and $\tilde R$, with the corrected actions
\begin{align}
    \mathcal{F}(\mu)&=\tilde F_{a}\wedge\mu^a\:,\\
    (\mathcal{F}^*(\alpha))_a&=\tr(\tilde F_{a}\wedge\alpha)\:,\\
    (\tilde{R}\cdot\tilde{\nabla}\mu)_a&=-{\tilde R}_a{}^b{}_c\wedge\tilde\nabla^+_b\mu^c\:.
\end{align}
In the last equation, $\tilde\nabla^+$ acts as the Bismut connection of the corrected, in general non-Kähler, metric $\tilde g$ on any free holomorphic vector indices, and as the Chern connection on form indices, so that the corrected $\DD$ operator squares to zero~\cite{McOrist:2024zdz, deLazari:2024zkg}. The frame rotation $\gamma$ appearing in the anomalous phase \eqref{eq:AnomPhase2} is also corrected to ${\gamma^a}_b=\tilde\nabla^+_b v^a$.

The variation of the new counter-term \eqref{eq:CounterX2} then becomes
\begin{align}
    \delta S^2_{\rm counter}(x)&=\beta'\int_X\bigl(\tr(\tilde F\epsilon)-\tr(\tilde R\gamma)\bigr)\wedge\tfrac{1}{4!}\omega^4+2\ii\int_X {\cal L}_v\omega\wedge\tfrac{1}{4!}\omega^4\:.
\end{align}
The first term cancels the anomalous phase transformation \eqref{eq:AnomPhase2}, while the second term vanishes as
\begin{equation}
    \int_X{\cal L}_v(\omega)\wedge\tfrac{1}{4!}\omega^4=\int_X{\cal L}_v\left(\tfrac{1}{5!}\omega^5\right)=0\:,
\end{equation}
since the holomorphic Lie derivative of a top-form is $\dd$-exact. Using the anomalous phase \eqref{eq:AnomPhase2}, we have thus succeeded in constructing a gauge invariant one-loop theory using only local counter-terms. The one-loop action reads
\begin{equation}
    S_{\text{one-loop}}=S+S^2_{\rm counter}(x)\:,
\end{equation}
where again $S$ is the classical master action \eqref{eq:BVaction2}, with the new corrected $\DD$ operator. 

Let us now return to the deformed instanton equations, \eqref{eq:DefInst1}--\eqref{eq:DefInst2}, to argue that solutions exists. Given a solution $h$ of the background hermitian Yang--Mills equation
\begin{equation}
    \omega^5\wedge F(h)=0\:,
\end{equation}
we deform $h$ to $\tilde h=h+\delta h$. To first order in $\delta h$, the deformed hermitian Yang--Mills equation reduces to
\begin{equation}
    \Delta_h\delta h=\text{Fifth and higher order in curvatures.}
\end{equation}
That is, the correction to the hermitian bundle metric on $V$ should solve a certain linearised system with higher-order curvatures as a source term. For stable bundles without holomorphic sections, this equation has a unique solution. Assuming a large-volume background, we can proceed in this fashion to solve Equation \eqref{eq:DefInst1} perturbatively, order by order in the curvatures. Similar reasoning holds for \eqref{eq:DefInst2}, where the background starting point can be taken to be the curvature of the Ricci-flat Calabi--Yau metric.

\subsection{Non-global cancellation}
\label{sec:NonGlobal}
Finally, via an alternative descent procedure, which again comes down to the choice of phase convention for the partition function,  
the anomalous phase transformation of the partition function may instead be written as
\begin{equation}
\label{eq:AnomalousPhase3}
    \delta\log(Z)=\ii\beta'\int_X\left(\tr(R^2)-\tr(F^2)\right)\wedge{\cal P}(\gamma,\epsilon)_{(3,3)}\:,
\end{equation}
where ${\cal P}(\gamma,\epsilon)_{(3,3)}$ is a $(3,3)$-form constructed from the curvatures and the gauge parameters $\gamma$ and $\epsilon$. Locally, we will have 
\begin{equation}
    \tr(R^2)=\partial\omega_{\text{CS}}(\nabla)_{(1,2)}\:,\quad\tr(F^2)=\partial\omega_{\text{CS}}(A)_{(1,2)}\:,
\end{equation}
for some Chern--Simons $(1,2)$-forms which we can take to be gauge invariant and $\delb$-closed.

If we then view the $T^{1,0}$-valued part of $y^{0,3}$ as a $(4,3)$-form $\chi$ (using the holomorphic top-form $\Omega$) we can cancel the anomaly \eqref{eq:AnomalousPhase3} by adding the non-global counter-term
\begin{equation}
\label{eq:counterChi}
    S_{{\rm counter}}(\chi)=\int_X\bigl(\omega_{{\rm CS}}(A)_{(1,2)}-\omega_{{\rm CS}}(\nabla)_{(1,2)}\bigr)\wedge\chi
\end{equation}
to the action and then allowing an anomalous gauge transformation for $\chi$:
\begin{equation}
\label{eq:GSchi}
    \delta\chi=\beta'\,\partial{\cal P}(\gamma,\epsilon)_{(3,3)}\:.
\end{equation}
Note that this transformation of $\chi$ resembles the Green--Schwarz-like transformations introduced when cancelling anomalies in holomorphic gauge theories coupled to gravity (see for example \cite{Costello:2015xsa, Williams:2018ows, Costello:2019jsy, Costello:2021bah, Costello:2021kiv, Costello:2022wso, Bittleston:2022nfr, Bittleston:2024efo}). Note also that $S_{\text{counter}}(\chi)$ is invariant under the classical $\delb$-exact transformation of $\chi$.

As noted above, the classical action remains gauge invariant modulo higher-loop effects, though it would be interesting to attempt to construct an exactly gauge-invariant one-loop effective action, as we did for the anomaly given in \eqref{eq:AnomalousPhase} in the previous subsections. This is expected to produce interesting geometric structures that are in a certain sense dual to the operator \eqref{eq:barD} and its associated structures. We will leave this to future work.  

Note also that the counter-term \eqref{eq:counterChi} is only locally well defined in general. A more thorough treatment of the anomaly would require extending the theory to some eleven manifold which is bounded by the five-fold $X$, as in the Green--Schwarz mechanism in, for example, six-dimensional supergravity~\cite{Monnier:2018nfs}. Such considerations can have interesting consequences for global anomalies; for local anomalies, the counter-term in \eqref{eq:counterChi} is sufficient. 

We can say more in the case where ${\rm ch}_2(X)$ and ${\rm \ch}_2(V)$ are equal in cohomology, such as the Hull--Strominger system. The squared curvatures are equal up to a $\partial$-exact term:
\begin{equation}
    \tr(R^2)-\tr(F^2)=\del T\:,
\end{equation}
for some $\delb$-closed $(1,2)$-form $T$,\footnote{In fact, on a K\"ahler background, $T$ is also $\delb$-exact by the $\del\delb$-lemma.} so the anomalous transformation becomes
\begin{equation}
    \delta\log(Z)=\ii\beta'\int_X\del T\wedge{\cal P}(\gamma,\epsilon)_{(3,3)}\:.
\end{equation}
In this case, we can use a local counter-term
\begin{equation}
    S_{{\rm counter}}(\chi)=-\int_XT\wedge\chi\:,
\end{equation}
with the anomalous gauge transformation as in \eqref{eq:GSchi} to cancel the anomaly. 

\subsection{Comparing the methods of anomaly cancellation}
\label{sec:Comp}
We have summarised our different approaches for cancelling the anomaly in Table \ref{tab:CancMethods}. As noted above, an anomaly is present when the phase of the partition function is not a well-defined function on the parameter space. Once it is established that the anomaly can be cancelled, and a global phase of the partition function may be defined, there are infinitely many ways to cancel the anomaly, corresponding to different conventions for the phase. Which convention one chooses is physically irrelevant, being related to different regularisation schemes in the field theory. In the BV approach to quantisation, the convention is fixed by choices made in the descent procedure and which connections one uses when defining the anomaly polynomials.

\begin{table}
{\small
\begin{tabularx}{1\textwidth}{>{\raggedright\arraybackslash}X>{\raggedright\arraybackslash}X>{\raggedright\arraybackslash}X}
\toprule 
\textbf{Cancellation method}  & \textbf{Advantages}  & \textbf{Disadvantages} \tabularnewline
\midrule
\midrule 
Non-local  & Simple, no need to adjust kinetic term in effective theory  & Non-local one-loop effective theory\tabularnewline
\addlinespace
Quasi-local  & Resembles usual Green--Schwarz mechanism  & Mildly non-local effective theory with standard hermitian Yang--Mills
and Levi--Civita curvatures in anomaly polynomial\tabularnewline
\addlinespace
Local  & Local one-loop effective theory & Need deformed instantons -- curvatures no longer solve standard Hull--Strominger
instanton condition\tabularnewline
\addlinespace
Non-global  & Uses Green--Schwarz mechanism of Costello--Li  & Non-global generic effective theory -- approach not fully developed
in current paper \tabularnewline
\bottomrule
\addlinespace
\end{tabularx}
}
\caption{A comparison of the different anomaly cancellation methods considered in the current paper, together with their advantages and disadvantages.}
\label{tab:CancMethods}
\end{table}

However, it is standard quantum field theory lore that terms in the Lagrangian should be local, including counter-terms in the effective theory, rendering the first two methods we present for cancelling the anomaly somewhat unsatisfactory. The third method renders the one-loop effective action fully local. However, we needed to adjust the connections used in the anomaly polynomial away from the standard hermitian Yang--Mills and Levi--Civita connections.\footnote{Cancelling the anomaly using local counter-terms and standard curvatures in the anomaly polynomial seems difficult, though not for lack of trying on the part of the authors.}  It is however interesting to note that the deformed, or quantum corrected, instanton equations \eqref{eq:DefInst1}-\eqref{eq:DefInst2} have terms of similar form to higher $\alpha'$ corrections of the BPS equations in heterotic supergravity \cite{Bergshoeff:1989de}, and investigating this connection further is part of future studies. 

Our final method of anomaly cancellation, studied in Section \ref{sec:NonGlobal}, employed a different descent procedure, leading to a Green--Schwarz mechanism akin to that employed by Costello--Li \cite{Costello:2015xsa}, where the Beltrami differential $\mu$ has an anomalous transformation. We also allowed for in principle non-globally well-defined counter-terms.  We did not attempt to construct a fully gauge invariant one-loop action in this case, which would be interesting and is part of future work. Indeed, it is expected that doing so will lead to an interesting new geometric structure, physically dual to the operator $\DD$, as the structure arises via employing a different but physically equivalent method to cancel the anomaly.

\section{Discussion and outlook}\label{sec:discussion}

In this paper, we have studied a ten-dimensional version of the six-dimensional superpotential theory of \cite{Ashmore:2018ybe}. We computed the one-loop partition function in terms of holomorphic Ray--Singer torsions, and studied the corresponding anomaly. We find an anomaly polynomial mimicking that of Green and Schwarz for ten-dimensional supergravity coupled to Yang--Mills \cite{Green:1984sg}, supporting the conjecture that our theory is the twisted version of this theory. Cancelling the anomaly via a Green--Schwarz mechanism, we also find a gauge invariant one-loop effective action if we promote the differential of the theory to $\DD$, a recently discovered operator whose first cohomology counts the infinitesimal deformations of heterotic solutions modulo ${\cal O}(\alpha'^2)$ corrections \cite{McOrist:2021dnd}. 
Whereas the theory constructed in Sections~\ref{sec:10d-theory} and~\ref{sec:cubic_theory} should be thought of as the Kodaira-Spencer theory corresponding to two-derivative classical supergravity, we can think of the anomaly-free theory constructed in Section~\ref{sec:anomalies} by adding local counterterms as the linearised part of the full heterotic Kodaira-Spencer theory to ${\cal O}(\alpha')$. These are our main results.

One way to understand our result for the anomaly is to note that the Dolbeault complex $(\Omega^{0,\bullet}, \bar\der)$ is spinor type, in the nomenclature of~\cite{Kupka:2024rvl}: we have that under the $\SU(5)$ structure group, the positive and negative chirality spinor bundles $S^{\pm}$ are isomorphic to $\Lambda^{0,\text{even}}T^*$ and $\Lambda^{0,\text{odd}}T^*$, and under this identification the Dirac operator becomes a linear combination of $\bar\der$ and $\bar\der^\dagger$. Also, the bundle $Q$ is isomorphic to $TX \oplus \End(V)$, so that overall the fields in $\Omega^{0,\text{even}}(Q)$ becomes the $\SU(5)$ decompositions of gravitino and gaugino fields. In fact, the BV complex of Figure \ref{fig:10d-Q-complex} becomes equivalent to the fermionic fields (plus anti-fields) of the supergravity theory, plus the ghosts and anti-ghosts of the local supersymmetry. Once this identification of the field content is made, it is perhaps not so surprising that the anomaly comes out to match the supergravity anomaly, which is entirely due to the presence of the chiral fermionic fields. 
In fact, under the $\SU(5)$ decomposition, the bosonic fields take the same form as the fermions, so that the field content of our theory represents exactly half of the total supergravity degrees of freedom. This is again what you might naively expect to emerge from a supersymmetric twist.

While this provides evidence that our theory is the twist of supergravity, one would like to derive this result systematically from a BV formulation of the supergravity theory in ten dimensions. 
Recently, using the framework of generalised geometry, the ten-dimensional supergravity action, including higher-order fermion terms, was written down in a very natural and compact way \cite{Kupka:2024tic, Kupka:2024vrd, Kupka:2024xur} (see also~\cite{Baron:2024tph}), which facilitated the formulation of a BV master action for the theory~\cite{Kupka:2025hln} in the component field formalism. This construction could therefore be used as the starting point for a full proof of our conjecture. 

However, in doing so, one would need to make explicit the relationship between the degrees of freedom in Figure~\ref{fig:10d-Q-complex} and corresponding quantities in generalised geometry. 
In~\cite{Kupka:2024rvl}, the corresponding complex in six dimensions was formulated in terms of generalised geometry variables. There, a bosonic field configuration and supersymmetry parameter are encoded as an $\SU(3) \times \SO(6+n)$ structure on the generalised tangent bundle in $\SO(6,6+n)\times \bbR^+$ generalised geometry, as described in~\cite{Ashmore:2019rkx}. Supersymmetry of such a configuration then corresponds to the vanishing of its generalised intrinsic torsion (as defined in general in~\cite{CSW4}). One can also consider a weaker $U(3) \times \SO(6+n)\times\bbR^+$ structure, analogous to requiring a K\"ahler rather than a Calabi--Yau metric in ordinary complex geometry. Associated to this weaker structure is a double complex, the BPS complex~\cite{Kupka:2024rvl}, which plays the role of the Dolbeault complex in generalised geometry. This object is the analogue of Figure \ref{fig:10d-Q-complex-BRST} in that setting. 
A natural extension of this complex 
possesses a BV symplectic pairing, exactly as in Section~\ref{sec:10d-BV-complex}, providing the analogue of the complex of Figure \ref{fig:10d-Q-complex} and indeed the resulting field theory is the linearised superpotential theory. As they are describing the same linearised theory and have isomorphic vector spaces, the complex derived from generalised geometry must be related to the one described in~\cite{Ashmore:2023vji} by a field redefinition. 
In the ten-dimensional scenario, one can instead consider a $U(5) \times \SO(10+n)\times\bbR^+$ structure in $SO(10,10+n)\times\bbR^+$ generalised geometry. This also has an associated BPS complex which one can extend in the same way to obtain one equipped with a BV symplectic pairing. The field theory associated to this should recover the quadratic part of~\eqref{eq:BVaction0} via a similar field redefinition as in the six-dimensional case. 
Armed with this field redefinition, one should then be able to relate the theory constructed here directly to that which arises from carrying out the Costello--Li twist of the BV action of supergravity in the generalised geometry description.

There are many interesting implications of our results, both from a mathematical and physical perspective. Mathematically, the operator $\DD$ is not the $(0,1)$-part of a connection, though it does define $Q$ as a double extension of bundles, but with non-tensorial extension classes. Locally on a patch ${\cal U}_i$, the operator $\DD$ has been shown to be trivialisable \cite{deLazari:2024zkg}
\begin{equation}
    \DD=G_i^{-1}\circ\delb\circ G_i\:,
\end{equation}
but where the ``local gauge transformations" $G_i$ are now linear operators rather than matrix-valued functions. Still, in this local frame one can now use transition operators 
\begin{equation}
    \psi_{ij}=G_i\circ G_j^{-1}
\end{equation}
to patch $Q$ together as a holomorphic sheaf. The transition operators are holomorphic in the sense that they commute with $\delb$
\begin{equation}
    [\delb,\psi_{ij}]=0\:.
\end{equation}
This local frame also allows one to connect the geometric structure to a more algebraic framework via Čech cohomology and algebraic geometry. In particular, a version of the Dolbeault theorem for such structures has been proven in \cite{deLazari:2024zkg, Ibarra2025}. Given the importance of heterotic theory for connecting string theory to realistic models of the physical world, a further mathematical investigation and classification of such geometric structures is warranted. Especially as the appearance of this type of more general extension structures also seems ubiquitous in higher derivative supergravity. For example, preliminary investigations of the corrected kinetic operator of the dual anomaly cancellation procedure of Section \ref{sec:NonGlobal} show a similar geometric structure emerging. 

There are many other directions for further investigation. For example, the anomalous dependence of the one-loop partition function on the background metric was studied in \cite{Ashmore:2023vji} for the six-dimensional theory. It was found that, given certain topological constraints, suitable counter-terms may be added to cancel these geometric anomalies, resulting in a topological invariant. A natural next step is to consider similar geometric dependencies of the partition function in the ten-dimensional case and perhaps derive a holomorphic anomaly equation. In a different direction, to apply the results of \cite{bismut1988analytic1,bismut1988analytic2,bismut1988analytic3}, one must assume a K\"ahler background. There are many truly non-K\"ahler solutions to the Hull--Strominger system, such as principal $\text{T}^2$ fibrations over K3 manifolds (see for example \cite{Dasgupta_1999,Goldstein_2004,fu2008theory,Becker:2009df,Melnikov:2014ywa,GarciaFernandez2020} and references therein). One could study anomalies of the theory for such backgrounds and adapt the results of \cite{bismut1988analytic1,bismut1988analytic2,bismut1988analytic3} to torsional geometries. In another direction, an open question is whether the particular deformed instanton equations introduced in Section~\ref{sec:deformed_instantons} arise from $\alpha'$ corrections in supergravity, or whether the ability to find appropriate counter-terms is related to $\Pi$-stability~\cite{Douglas:2000ah} of the background.

Finally, it would be interesting to extend the quantum corrected quadratic action, with the corrected operator $\DD$, to a proper non-linear theory \`a la the classical theory \eqref{eq:CubicSheafTheory}. This is closely related to understanding the full $L_\infty$ structure of the Hull--Strominger system, similar to what is done in \cite{Costello:2021kiv} for type I BCOV theory on torsion-free zero flux backgrounds, and on $\mathbb{C}^5$ in particular. In this regard, one might also wonder if the theory has anomalies at higher loop order. Indeed, the theory we consider is power-counting non-renormalizable. This means that gauge anomalies (and counter-terms) are not restricted to one loop and can appear at any order in the loop expansion. In previous examples, cancelling the one-loop anomalies has been enough to ensure cancellation to all loop orders~\cite{Costello:2012cy, Costello:2015xsa, Bittleston:2022nfr}. Given the close relation between our theory and type I BCOV \cite{Costello:2015xsa}, as described in Section \ref{sec:cubic_theory}, we are hopeful that the same will be true here. 

\acknowledgments
We thank Alex S.\,Arvanitakis, Xenia de la Ossa, Mario Garcia-Fernandez, Pedram Hekmati, Dimitri Kanakaris, Julian Kupka, Magdalena Larfors, Hannah de L\'azari, Jason D. Lotay, Matthew Magill, Jock McOrist, Sebastien Picard, Henrique S\'a Earp, Ingmar Saberi, Martin Sticka, David Tennyson and Fridrich Valach for interesting and insightful conversations related to the work presented here. C.S.-C.~is supported by an EPSRC New Investigator Award, grant number EP/X014959/1. No new data was collected or generated during the course of this research.

\bibliographystyle{JHEP}
\bibliography{citations}

\end{document}